\documentclass[a4paper,12pt]{article}
\pdfoutput=1

\usepackage[a4paper,left=80pt,right=80pt,top=85pt,bottom=97pt]{geometry}
\usepackage[latin1]{inputenc}
\usepackage{dsfont}
\usepackage{amsfonts,amsmath,amssymb}
\usepackage{amsthm}
\usepackage{graphicx}
\usepackage[small,bf]{caption}
\usepackage{subcaption}
\usepackage{booktabs}
\usepackage[numbers, sort]{natbib}
\allowdisplaybreaks[1]
\usepackage{color}
\usepackage{ulem}
\usepackage{tikz}
\usepackage{multirow}
\usetikzlibrary{arrows}
\usepackage{xspace}
\def\bal#1\eal{\begin{align}#1\end{align}}
\newcommand{\be}{\begin{equation}}
\newcommand{\ee}{\end{equation}}
\newcommand{\bea}{\begin{eqnarray}}
\newcommand{\eea}{\end{eqnarray}}
\newcommand{\besub}{\begin{subequations}}
\newcommand{\eesub}{\end{subequations}}
\newcommand{\ba}{\begin{array}}
\newcommand{\ea}{\end{array}}
\newcommand{\bi}{\begin{itemize}}
\newcommand{\ei}{\end{itemize}}
\newcommand{\nn}{\nonumber}

\newcommand{\Mcal}{{\cal M}}
\newcommand{\Ocal}{\ensuremath{\mathcal O}}
\newcommand{\Lcal}{{\cal L}}

\newcommand{\hc}{{\textrm{h.c.}}}

\begin{document}

\begin{titlepage}

\flushright{ 
 LPT-Orsay-15-35
 \\
 HIP-2015-19/TH}
 
\vspace*{2.5cm}

\begin{center}
{\Large 
{\bf
Non-Abelian gauge fields as dark matter
}}
\\
[1.5cm]

{
{\bf
Christian Gross$^{1}$, Oleg Lebedev$^{1}$, Yann Mambrini$^{2}$
}}
\end{center}
%\addtocounter{footnote}{-4}

\vspace*{0.5cm}

\centering{
$^{1}$ 
\it{Department of Physics and Helsinki Institute of Physics, \\
Gustaf H\"allstr\"omin katu 2, FI-00014 Helsinki, Finland
}

\vspace*{0.15cm}
$^{2}$ 
\it{Laboratoire de Physique Th\'eorique, CNRS -- UMR 8627, \\
Universit\'e de Paris-Sud 11, F-91405 Orsay Cedex, France
}
}

\vspace*{0.4cm}

\vspace*{1.2cm}

\begin{abstract}
\noindent
SU(N) Lie algebras possess discrete symmetries which can lead naturally to stable vector dark matter (DM). In this work, we consider the possibility that the dark SU(N) sector couples to the visible sector through the Higgs portal. We find that minimal {\it CP}--conserving hidden ``Higgs sectors'' entail stable massive gauge fields which fall into the WIMP category of dark matter candidates. For SU(N), $N>2$, DM consists of three components, two of which are degenerate in mass. In all of the cases, there are substantial regions of parameter space where the direct and indirect detection as well as relic abundance constraints are satisfied. 
\end{abstract}

\end{titlepage}
\newpage

\tableofcontents

%=========================================================================
%=========================================================================
\section{Introduction}
%=========================================================================
%=========================================================================
The puzzle of dark matter remains an outstanding problem of particle physics. One of the more attractive approaches to this problem exploits the fact that weakly interacting massive particles (WIMPs) in thermal equilibrium produce the relic dark matter abundance in the right ballpark. Particles of this type appear in many extensions of the Standard Model. In this work, we explore the possibility that the Standard Model is connected to the dark sector through the Higgs portal~\cite{Silveira:1985rk},
\begin{equation}
V_{\rm portal}= \lambda_{h \phi} H^\dagger H \phi^\dagger \phi \;,
\label{eq1}
\end{equation}
where $H$ is the Higgs field and $\phi$ is the ``hidden Higgs'', that is, the field responsible for breaking the gauge symmetry of the hidden sector. We assume the dark sector to be endowed with U(1) or SU(N) gauge symmetry. In that case, the massive gauge fields can play the role of WIMP--type dark matter quite naturally.
Indeed, they are weakly coupled to the Standard Model and, owing to inherent discrete symmetries, can be stable.

The U(1) case was considered in Ref.~\cite{Lebedev:2011iq} where the stabilising $Z_2$ symmetry was found to be related to charge conjugation.
The SU(2) example was worked out in Ref.~\cite{Hambye:2008bq} and the DM stability was attributed to a custodial SO(3). In this work, we extend these ideas to larger SU(N) Lie groups and uncover the nature of the underlying discrete symmetries. We focus on weakly coupled theories, although
confining hidden sectors represent a viable alternative~\cite{Hambye:2009fg,Boddy:2014yra}.
Finally, we perform a comprehensive study of DM phenomenology in all of these cases, which includes direct and indirect DM detection as well as an analysis of the DM relic abundance. Our main conclusion is that massive hidden gauge fields serve as viable and attractive DM candidates.

%=========================================================================
%=========================================================================
\section{Massive U(1) and SU(2) gauge fields as vector DM}
%=========================================================================
%=========================================================================
In this section, we review the cases of massive U(1) and SU(2) gauge fields as vector DM and identify the underlying symmetries leading to stability of vector dark matter.
In what follows we assume that the hidden sector consists only of gauge fields and of scalar multiplets which are necessary to make these gauge fields massive.

%=========================================================================
\subsection{Hidden U(1)}
%=========================================================================
An Abelian gauge sector provides the simplest example of the vector DM model endowed
with a natural $Z_2$ symmetry~\cite{Lebedev:2011iq}. In this case, the $Z_2$ corresponds to the charge conjugation symmetry.
 
 Consider a U(1) gauge theory with a single charged scalar $\phi$,
 \begin{equation}
 {\cal L_{\rm hidden}}= -{1\over 4} F_{\mu\nu} F^{ \mu\nu} + (D_\mu \phi)^\dagger D^\mu \phi -V(\phi) \;,
 \end{equation}
 where $F_{\mu\nu}$ is the field strength tensor of the gauge field $A_\mu$ and 
 $V(\phi)$ is the scalar potential. For easier comparison to the non--Abelian case, we take the charge of $\phi $ to be +1/2. Suppose at the minimum of the scalar potential
 $\phi$ develops a VEV, $\langle \phi \rangle= 1/\sqrt{2}~ \tilde v$ . The imaginary part of $\phi$ gets eaten by the gauge field which
 now acquires the mass $m_A= \tilde g \tilde v /2$, where $\tilde g$ is the gauge coupling.
 The real part of $\phi$ remains as a degree of freedom. Denoting it by $\rho$ and normalising it canonically, $\phi=1/\sqrt{2}~ (\rho +\tilde v)$ , we get the following gauge--scalar interactions:
 \begin{eqnarray}
&& \Delta {\cal L}_{\rm s-g}= {\tilde g^2\over 4} \tilde v \rho \; A_\mu A^{ \mu} +
 {\tilde g^2\over 8} \rho^2 \; A_\mu A^{\mu} \;. 
 \end{eqnarray}
The system possesses the $Z_2$ symmetry 
\begin{equation}
A_\mu \rightarrow - A_\mu ~,
\end{equation} 
which is the usual charge conjugation symmetry. In terms of the original scalar field, this symmetry acts as $\phi \rightarrow \phi^*$ and $A_\mu \rightarrow - A_\mu$, which is preserved by both the Lagrangian and the vacuum. The $Z_2$ makes the massive gauge field stable.
Note that this symmetry applies to a sequestered U(1) which has no tree
level mixing with the hypercharge, in which case no mixing is generated radiatively either.

Interactions with the visible sector proceed through the Higgs portal coupling
\begin{equation}
 {\cal L_{\rm portal}}= - \lambda_{h \phi} \vert H \vert^2 \vert \phi \vert^2 \;,
 \end{equation}
which also leads to the Higgs mixing with $\rho$. 
In the unitary gauge, the Higgs field is given by $H^T= (0, v+h)/\sqrt{2}$.
The fields $\rho$ and $h$ are then to be expressed in terms of the mass eigenstates $h_{1,2}$ as follows: 
\begin{eqnarray}
&& \rho= - h_1 \; \sin\theta + h_2 \; \cos\theta \;, \nonumber\\
&& h = h_1 \; \cos\theta + h_2 \; \sin\theta \;,
 \end{eqnarray}
where the mixing angle $\theta$ is constrained by various experiments, most notably, by LEP and LHC. 
The upper bound on $\sin \theta$ depends on the mass of the heavier state $h_2$ and is around 0.3 for $m_{h_2}$ of the order of a TeV, see e.g.~figure 3 of Ref.~\cite{Falkowski:2015iwa} for details.
The lighter state $h_1$ is identified with the 125 GeV Higgs, while the mass of the second state can vary in a wide range.

As the Higgs portal necessarily preserves the $Z_2$ symmetry, $A_\mu$ 
is a viable DM candidate.
All the relevant scattering processes proceed through an exchange of $h_1$ and $h_2$, which include DM annihilation into the SM particles as well as DM scattering on nucleons.

%=========================================================================
\subsection{Hidden SU(2)}
%=========================================================================
The U(1) considerations can easily be extended to SU(2), albeit with a modification of the stabilising symmetry.
Consider an SU(2) gauge theory with one doublet $\phi$,
 \begin{equation}
 {\cal L_{\rm hidden}}= -{1\over 4} F_{\mu\nu}^a F^{a \; \mu\nu} + (D_\mu \phi)^\dagger D^\mu \phi -V(\phi) \;,
 \end{equation}
where $a=1,2,3$.
The potential is assumed to have a minimum at a nonzero VEV of $\phi$. In the unitary gauge, $\phi$ takes the form 
\begin{equation}
\phi = {1\over \sqrt{2}}\; \left( 
\begin{matrix}
 0 \\
 \rho +\tilde v
\end{matrix}
\right)~,
\end{equation}
with $\rho$ being a real field and $\tilde v$ being the VEV. Denoting the gauge coupling by $\tilde g$,
this leads to the gauge boson mass $m_A= \tilde g \tilde v/2$. The scalar--gauge and gauge--gauge field interactions are given by
\begin{eqnarray}
&& \Delta {\cal L}_{\rm s-g}= {\tilde g^2\over 4} \tilde v \rho \; A_\mu^a A^{ a\; \mu} +
 {\tilde g^2\over 8} \rho^2 \; A_\mu^a A^{ a\; \mu} \;, \nonumber \\
&& \Delta {\cal L}_{\rm g-g} = - \tilde g \epsilon^{abc} (\partial_\mu A_\nu^a) A^{\mu \;b}
A^{\nu \;c} -{\tilde g^2\over 4} \left( (A_\mu^a A^{\mu \;a})^2 -
 A_\mu^a A_\nu^a \; A^{\mu \;b} A^{\nu \;b} \right) ~. 
\end{eqnarray} 
Although the triple gauge vertex breaks the parity of the previous section,
it follows that the system possesses a $Z_2 \times Z_2'$ symmetry,
\begin{eqnarray}
&& Z_2: ~~ A^1_\mu \rightarrow - A^1_\mu ~~,~~ A^2_\mu \rightarrow - A^2_\mu \;, \nonumber \\
&& Z_2':~~ A^1_\mu \rightarrow - A^1_\mu ~~,~~ A^3_\mu \rightarrow - A^3_\mu \;.
\end{eqnarray}
As a result, all three $A^a_\mu$ fields are stable and can play the role of dark matter.
While the above symmetry is 
sufficient to ensure stability of DM, it actually generalises in this simple case to a custodial 
SO(3)~\cite{Hambye:2008bq}. As we will see, for larger SU(N) groups, it is the discrete
symmetry that plays a crucial role. The first $Z_2$ is associated with a gauge transformation, while the $Z_2'$ generalises the charge conjugation symmetry, i.e.~it
corresponds to complex conjugation of the group elements.

As before, the dark sector couples to the visible one through the Higgs portal,
$\lambda_{h \phi} \vert H \vert^2 \vert \phi \vert^2 $. The discussion of the Higgs mixing with $\rho$ of the previous section applies here as well.
Clearly, the $Z_2 \times Z_2'$ symmetry is preserved by the Higgs portal and the hidden gauge fields couple to the visible sector only in pairs.

%=========================================================================
%=========================================================================
\section{Extension to SU(3)}
%=========================================================================
%=========================================================================
While the SU(2) case is straightforward, larger SU(N) groups exhibit more complicated breaking patterns. In a phenomenologically viable set--up, the symmetry must be broken completely to avoid the existence of massless fields (barring confinement). One possibility 
would be to break SU(3) by two scalar multiplets in the fundamental representation, i.e.~triplets. One may also explore other options involving SU(3) tensors. As we detail below,
our conclusion is that a single irreducible representation with two indices cannot break SU(3) completely, which leaves the two triplet option as the minimal one.

%=========================================================================
\subsection{Breaking SU(3) by tensor fields}
%=========================================================================
The lowest order SU(3) tensor whose generic VEV can break SU(3) completely is the symmetric tensor $\phi_{ij}$, that is {\bf 6} of SU(3). 
Gauge transformations act on $\phi_{ij}$ as
\begin{equation}
\phi \rightarrow U \phi \, U^T \;.
\end{equation}
By virtue of Takagi's matrix decomposition, this allows one to bring $\phi_{ij}$
to the diagonal form, 
\begin{equation}
\phi = \left( 
\begin{matrix}
 \phi_1 & 0 & 0 \\
 0 & \phi_2 & 0 \\
 0 & 0 & \phi_3
\end{matrix}
\right)~,
\end{equation}
where $\phi_1, \phi_2, \phi_3$ are real up to an overall complex phase.
If the VEVs of $\phi_1, \phi_2 $ and $\phi_3$ are all different, SU(3) is broken completely. However,
when some of them coincide, the residual gauge group is at least SO(2).

In order to determine what VEVs are possible, let us write down the most general gauge invariant potential for $\phi_{ij}$.
It is easy to convince oneself that the potential has the form
\begin{equation}
V= m^2 \; {\rm Tr} \phi^\dagger \phi + \lambda_1 {\rm Tr} (\phi^\dagger \phi)^2
+ \lambda_2 \left( {\rm Tr} \phi^\dagger \phi \right)^2 +
 \left( \mu \; {\rm Det} \phi + {\rm h.c.} \right) \;,
\end{equation}
where $m^2$ can be negative.
The minima of this potential determine the $\phi_i$ VEVs.
Acting upon $V$ with the operator $\phi_i \partial/\partial \phi_i$ (no summation over $i$), one finds that all nonzero $ \langle \phi_i \rangle $ satisfy the same equation.
This implies that $ \langle \phi_i \rangle $ are either degenerate or zero. 
For the case $\mu=0$,
such behaviour has been noticed in~\cite{Li:1973mq,Jetzer:1983ij}.
We find (analytically and numerically) that it persists in the case of nonzero $\mu$ as well.
As a result, the residual gauge symmetry is at least SO(2) and the corresponding gauge bosons remain massless.\footnote{ Ref.~\cite{D'Eramo:2012rr} has considered $\langle \phi \rangle$ proportional to the unit matrix, which entails unbroken SO(3). The corresponding gauge bosons may be confined in glueballs.}
Thus, the model with a single symmetric tensor is unrealistic. 

One may also consider the possibility of breaking SU(3) by an antisymmetric tensor.
By virtue of Youla's decomposition, it can be gauge--transformed to the block--diagonal form 
\begin{equation}
\phi = \left( 
\begin{matrix}
 0 & \phi_1 & 0 \\
 -\phi_1 & 0 & 0 \\
 0 & 0 & 0
\end{matrix}
\right)~.
\end{equation} 
Therefore, the residual gauge symmetry is at least Sp(1)=SU(2), which can also be understood via the equivalence between the antisymmetric tensor and the (anti)funda\-mental representation. Again we find that the model is unrealistic.

Similarly, complete SU(3) breaking cannot be achieved by a VEV of an adjoint scalar.
In this case, the group rank is preserved and thus there are massless gauge fields.

The next simplest option is to combine the symmetric and antisymmetric tensors. This system would have 9 complex degrees of freedom, while two SU(3) triplets only have 6.
Thus SU(3) breaking with two triplets is minimal and sufficient for our purposes.

%=========================================================================
\subsection{Breaking SU(3) by triplets and $Z_2 \times Z_2'$ }
%=========================================================================
Misaligned VEVs of two triplets break SU(3) completely. This breaking pattern can be understood in stages: the first triplet VEV reduces the symmetry to SU(2), while the second breaks the remaining SU(2). VEVs misaligned in SU(3) space represent a generic situation, that is, they result from the minimisation of a general scalar potential consistent with SU(3) symmetry. Therefore, such a breaking pattern is phenomenologically viable. 

Before going into details, let us identify the Lie group discrete symmetries which eventually lead to DM stability. One way to find them is to analyse the SU(3) structure constants (using the usual Gell-Mann basis),
\begin{eqnarray}
&& f^{123}=1 \;, \nonumber\\
&& f^{147}=-f^{156}= f^{246}=f^{257}=f^{345}=-f^{367}={1\over 2} \;, \nonumber\\
&& f^{458}=f^{678}= {\sqrt{3}\over 2} \;.
\end{eqnarray}
Identifying the transformation properties of the generators with those of the gauge fields, we define the ``parity'' transformation as
\begin{equation}
A^a_\mu \rightarrow \eta (a) A^a_\mu \;.
\end{equation}
It is easy to see that the structure constants are invariant if the parities are
\begin{eqnarray}
 Z_2: ~~&&\eta (a) = -1 ~~{\rm for }~~a=1,2,4,5 ~, \nonumber \\
 &&\eta (a) = +1 ~~{\rm for }~~a=3,6,7,8 ~,
\end{eqnarray}
and also
\begin{eqnarray}
 Z_2': ~~&&\eta (a) = -1 ~~{\rm for }~~a=1,3,4,6,8 ~, \nonumber \\
 &&\eta (a) = +1 ~~{\rm for }~~a=2,5,7 ~.
\end{eqnarray}
One may notice that the first $Z_2$ acts on the off--diagonal generators with nonzero entries in the first row, while the second reflects the real SU(3) generators.
In Section \ref{SUN}, we will show that these symmetries generalise to arbitrary SU(N) and 
that the first $Z_2$ is a gauge transformation, whereas the second corresponds to an outer automorphism of the group, i.e.~complex conjugation of the group elements.

These symmetries are inherited by the Yang--Mills Lagrangian. If {\it CP} is conserved, they are also preserved in the matter sector leading to stable dark matter. Below we study the relevant interactions in detail.

%=========================================================================
\subsection{Explicit example}
%=========================================================================
Consider an SM extension by two complex fields $\phi_1$ and $\phi_2$ transforming as triplets of hidden gauge SU(3).
The Lagrangian of the model is
\be
\Lcal_{\rm SM} + \Lcal_{\rm portal} + \Lcal_{\rm hidden} \;,
\ee
where 
\besub
\bal
-\Lcal_{\rm SM} &\supset V_{\rm SM} =\frac{\lambda_{H}}{2} |H|^4+m_{H}^2 |H|^2 \;,
 \\
-\Lcal_{\rm portal} &= V_{\rm portal}= \lambda_{H11} \, |H|^2 |\phi_1|^2 + \lambda_{H22} \, |H|^2 | \phi_2|^2 - ( \lambda_{H12} \, |H|^2 \phi_1^\dagger \phi_2 +\hc)\;,
 \\
\Lcal_{\rm hidden} &= - \frac12 \textrm{tr} \{G_{\mu \nu} G^{\mu \nu}\} + |D_\mu \phi_1|^2 + |D_\mu \phi_2|^2 -V_{\rm hidden} \,.
\eal
\eesub
Here, $G_{\mu \nu}=\partial_\mu A_\nu - \partial_\nu A_\mu + i \tilde g [A_\mu,A_\nu]$ is the field strength tensor of the SU(3) gauge fields $A_\mu^a$, $D_\mu \phi_i = \partial_\mu \phi_i + i \tilde g A_{\mu} \phi_i$ is the covariant derivative of $\phi_i$, $H$ is the Higgs doublet, and the most general renormalisable hidden sector scalar potential is given by 
\bal
V_{\rm hidden}(\phi_1,\phi_2) &=
m_{11}^2 |\phi_1|^2
+ m_{22}^2 |\phi_2|^2
- ( m_{12}^2 \phi_1^\dagger \phi_2 + \hc )
\nn \\ 
& 
+ \frac{\lambda_1}{2} |\phi_1|^4
+ \frac{\lambda_2}{2} |\phi_2|^4
+ \lambda_3 |\phi_1|^2 |\phi_2|^2
+ \lambda_4 | \phi_1^\dagger\phi_2 |^2
\nn \\ 
& 
+ \left[
\frac{ \lambda_5}{2} ( \phi_1^\dagger\phi_2 )^2
+ \lambda_6 |\phi_1|^2
( \phi_1^\dagger\phi_2)
+ \lambda_7 |\phi_2|^2
( \phi_1^\dagger\phi_2 )
+ \hc \right] \,.
\eal
Using SU(3) gauge freedom, 5 real degrees of freedom of $\phi_1$ and 3 real degrees of freedom of $\phi_2$ can be removed. Therefore, 
in the unitary gauge $\phi_1$, $\phi_2$ read
\be \label{unitarygauge}
\phi_1={1\over \sqrt{2}} \,
\left( \begin{array}{c}
0\\0\\v_1+\varphi_1
\end{array} \right) \,,
\quad 
\phi_2= {1 \over \sqrt{2}}\,
\left( \begin{array}{c}
0\\v_2+\varphi_2\\(v_3+\varphi_3) + i (v_4+\varphi_4)
\end{array} \right) ~,
\ee
where the $v_i$ are real VEVs and $\varphi_{1-4} $ are real scalar fields.
Analogously, we express the Higgs field in the unitary gauge as $H^T= (0,v+h)/\sqrt{2}$.

In what follows, we make two assumptions which are crucial for stability of vector dark matter:
\bi
\item
the scalar potential is {\it CP} invariant 
\item
the VEVs of $\phi_1$, $\phi_2$ are real.
\ei
The first condition implies that the scalar couplings are real, while the second assumes
that no spontaneous {\it CP} violation occurs ($v_4=0$). As a result, {\it CP}--even and {\it CP}--odd fields do not mix.

In this case, the $Z_2 \times Z_2'$ symmetry extends to the Higgs sector as well. 
Under the first $Z_2$ all $\varphi_i$ are even, while the second $Z_2'$ 
 reflects $\varphi_4$ and leaves the other fields intact. As we detail in Section \ref{SUN},
this assignment follows from the explicit form of the first $Z_2$ as a gauge transformation and the fact that the second $Z_2'$ acts as complex conjugation. In any case, these are explicit symmetries of the Lagrangian and the vacuum. 
The full list of the parities is presented in Table~\ref{parities}.
Clearly, the lightest states with non--trivial parities cannot decay to the Standard Model particles.
 \begin{table}[h]
 \begin{center}
 \begin{tabular}{|c|c|}
\hline
 fields & $Z_2 \times Z_2'$
 \\ \hline \hline 
$h, \varphi^1, \varphi^2, \varphi^3, A_\mu^7$ & $(+,+)$
\\ \hline 
$A_\mu^2,A_\mu^5$& $(-,+)$
\\ \hline 
$A_\mu^1,A_\mu^4$& $(-,-)$
\\ \hline 
$\varphi^4,A_\mu^3,A_\mu^6, A_\mu^8$& $(+,-)$
\\ \hline
\end{tabular}
\end{center}
\caption{\label{parities} $Z_2 \times Z_2'$ parities of the scalars and dark gauge bosons.
}
 \end{table}

We now discuss the Lagrangian in more detail, starting with the covariant derivates of $\phi_1$ and $\phi_2$,
\be \label{dphi}
|D_\mu \phi_i|^2 = |\partial_\mu \phi_i|^2 + i \tilde g A_\mu^a \left((\partial^\mu \phi_i)^\dagger T^a \phi_i - \hc \right)
+\tilde g^2 A_\mu^a A^{\mu b} \, \phi_i^\dagger T^a T^b \phi_i \,.
\ee
Inserting the parametrization~(\ref{unitarygauge}) with $v_4=0$,
we get the kinetic terms for the scalars
\be 
|\partial_\mu \phi_1|^2+ |\partial_\mu \phi_2|^2 
=
\frac12
\sum_{i=1}^4
(\partial_\mu \varphi_i)^2 \,,
\ee
the mass terms for the gauge fields, the mixing terms 
as well as the gauge--scalar interactions.
Let us first discuss the terms quadratic in the fields.
The third term on the r.h.s. of Eq.~(\ref{dphi}) contains the mass terms for the gauge bosons, 
\be
\Lcal \supset
\frac12 \left( \vec{A}_{(1,4)}^T \Mcal_{(1,4)} \vec{A}_{(1,4)}
+ \vec{A}_{(2,5)}^T \Mcal_{(2,5)} \vec{A}_{(2,5)}
+ \vec{A}_{(3,6,8)}^T \Mcal_{(3,6,8)} \vec{A}_{(3,6,8)}
+ \Mcal_{(7)} A_\mu^7 A^{\mu 7} \right),
\ee
where we have used the shorthand notation 
$\vec{A}_{(1,4)}^T \equiv (A_\mu^1,A_\mu^4)$ and similarly for the other gauge bosons.
The mass--squared matrices are
\bal \label{Amasses}
\Mcal_{(1,4)}&=\Mcal_{(2,5)}
= \frac{\tilde g^2}{4}
\left(
\begin{array}{cc}
v_2^2&v_2 v_3
\\
v_2 v_3&v_1^2+v_3^2
\end{array}
\right)~,
\nn \\
\Mcal_{(3,6,8)}
&= \frac{\tilde g^2}{4}
\left(
\begin{array}{ccc}
v_2^2&-v_2 v_3&-v_2^2/\sqrt{3}
\\
-v_2 v_3&v_1^2+v_2^2+v_3^2&-v_2 v_3/\sqrt{3}
\\
-v_2^2/\sqrt{3}&-v_2 v_3/\sqrt{3}&(4 v_1^2+v_2^2+4 v_3^2)/3
\end{array}
\right)~,
\nn \\
\Mcal_{(7)}
&= \frac{\tilde g^2}{4} ( v_1^2+v_2^2+v_3^2 ) \,.
\eal
The second term on the r.h.s. of Eq.~(\ref{dphi}) contains the $A_\mu \partial^\mu \varphi$ mixing terms
\be
\Lcal \supset \frac{\tilde g v_2}{2} A^{\mu 6} \partial_\mu \varphi_4
-\frac{\tilde g v_3}{\sqrt{3}} A^{\mu 8} \partial_\mu \varphi_4
+ \frac{\tilde g v_3}{2} A^{\mu 7} \partial_\mu \varphi_2
- \frac{\tilde g v_2}{2} A^{\mu 7} \partial_\mu \varphi_3 \,.
\ee
In general, terms of the type $\kappa_{ai} A_\mu^a \partial^\mu \varphi^i$ can be removed by the field redefinition 
\be \label{removemix}
\tilde A_\mu^a = A_\mu^a + \partial_\mu Y^a \;,
\quad \textrm{where} \quad
Y^a\equiv (\Mcal^{-1})_{ab}\, \kappa_{bi}\,\varphi^i\,.
\ee
We then have
\bal
&- \frac12 \textrm{tr} \{G_{\mu \nu} G^{\mu \nu}\} + \frac12 (\partial_\mu \varphi^i)^2 + \frac12 \Mcal_{ab} A_\mu^a A^{\mu b} + \kappa_{ai} A_\mu^a \partial^\mu \varphi^i =
\nn
\\
& - \frac12 \textrm{tr} \{\tilde G_{\mu \nu} \tilde G^{\mu \nu}\} + \frac12 \gamma_{ij} (\partial_\mu \varphi^i) (\partial^\mu \varphi^j) + \frac12 \Mcal_{ab} \tilde A_\mu^a \tilde A^{\mu b} \,,
\eal
where
\besub
\bal
\tilde G_{\mu \nu} &= \partial_\mu \tilde A_\nu - \partial_\nu \tilde A_\mu + i \tilde g [\tilde A_\mu - \partial_\mu Y,\tilde A_\nu - \partial_\nu Y] \, ,
\\
\gamma_{ij} &=\delta_{ij} - \kappa^T_{ia}\Mcal^{-1}_{ab}\kappa_{bj} \, ,
\eal
\eesub
and we have defined $Y=Y^a T^a$.
The kinetic terms for the vector fields are still canonically normalised, unlike those for the scalar fields.
The latter can be normalised canonically by a further field redefinition
\be \label{normalizescalars}
\tilde \varphi^i = \omega_{ik} \varphi^k \;, \quad \textrm{where} \quad (\omega^T \omega)_{ij}= \gamma_{ij} \,.
\ee
Note that the term $- \frac12 \textrm{tr} \{\tilde G_{\mu \nu} \tilde G^{\mu \nu}\} $ includes additional couplings of the gauge fields to the scalars.
The resulting vertices are obtained by 
 replacing $A_\mu^a$ in the triple and quartic gauge boson terms by $-\partial_\mu Y^a$
 such that interactions of the type 
 $(\tilde A)^3 \partial Y, (\tilde A)^2 (\partial Y)^2,$ etc. arise.
 The analysis of the general case is very cumbersome, so further we will focus on the simple case of $v_3 = 0$ which retains all the relevant physics.
 In that case, the couplings involving $\partial Y$ play no role in the DM phenomenology.

%=========================================================================
\subsection{Detailed study for $v_{3,4} = 0$}
%=========================================================================
As long as $v_1$ and $v_2$ are nonzero, SU(3) is broken completely. Hence it suffices
to consider the case $v_3 = 0$, which simplifies the analysis.
Then the only mixing term among the gauge bosons is $ A^3_\mu A^{\mu 8}$ and the only
gauge--scalar mixing terms are 
$ A^6_\mu \partial^\mu \varphi_4$, $ A^7_\mu \partial^\mu \varphi_3$.

In what follows, we restrict ourselves to the case $v_3 = 0$.

\subsubsection{Masses}

\paragraph{Gauge boson masses}

The gauge boson mass eigenstates are
\bal
\left( \begin{array}{c}
A'^3_\mu\\
A'^8_\mu
\end{array} \right)
=
 \left( \begin{array}{c}
\cos \alpha \, A^3_\mu + \sin \alpha \, A^8_\mu\\
\cos \alpha \, A^8_\mu - \sin \alpha \, A^3_\mu
\end{array} \right)\,,
\quad
\textrm{where}
\quad
\tan2 \alpha = \frac{\sqrt{3} v_2^2}{2 v_1^2-v_2^2} \;,
\eal
and the masses are\footnote{Note that
$\frac{\tan \alpha}{\sqrt{3}}=\frac{v_2^2}{4 v_1^2} + \Ocal\left(\frac{v_2^6}{v_1^6}\right)$,
$\cos \alpha = 1- \frac{3 v_2^4}{32 v_1^4} + \Ocal\left(\frac{v_2^8}{v_1^8}\right)$
and
$\sin \alpha = \frac{\sqrt{3} v_2^2}{4 v_1^2} + \Ocal\left(\frac{v_2^6}{v_1^6}\right)$.
}
\be
m^2_{A'^3}=\frac{\tilde g^2 v_2^2}{4} \Big(1-\frac{\tan \alpha}{\sqrt{3}}\Big) \,,
\qquad
m^2_{A'^8}=\frac{\tilde g^2 v_1^2}{3} \frac{1}{1-\frac{\tan \alpha}{\sqrt{3}}} \,,
\label{A3prime-mass}
\ee
while the other gauge boson masses can be read off directly from Eq.~(\ref{Amasses}) for $v_3=0$:
\be
m^2_{A^1}=m^2_{A^2}=\frac{\tilde g^2 }{4}v_2^2\;,
\quad
m^2_{A^4}=m^2_{A^5}=\frac{\tilde g^2 }{4}v_1^2\;,
\quad
m^2_{A^6}=m^2_{A^7}=\frac{\tilde g^2 }{4}(v_1^2+v_2^2) \;.
\ee
For $v_2< v_1$, the light fields are $A_\mu^{1,2}$ and $A_\mu^{\prime 3}$, with the latter being the lightest.
It is instructive
to consider the case $v_2^2 \ll v_1^2$, so that $\tan \alpha $ is small and positive.
Then $A'^3_\mu$ is slightly lighter than $A^1_\mu$ and $A^2_\mu$ by a factor of $(1-\frac{\tan \alpha}{\sqrt{3}})^{1/2}$, while the other five dark gauge bosons are all much heavier, by a factor of order $v_1/v_2$. The mass degeneracy between $A^1_\mu$ and $A^2_\mu$ persists at loop level by symmetry arguments (see Section~\ref{SUN}).

One can easily verify that the heavier states $A^{4-7}_\mu$ and $A^{\prime 8}_\mu$ 
all decay into the light states and the SM particles. The decay proceeds via emission
(off--shell or on--shell) of the {\it CP} even scalars which couple to the SM Higgs
 and thus to all other SM fields.

The 3 lightest states all have different $Z_2 \times Z_2^\prime$ parities such that they cannot decay into each other by emitting SM particles. The only scalar with negative parity is $\varphi_4$, however it is generally heavy (see below) and does not contribute to the above decay. Hence, $A_\mu^{1,2}$ and $A_\mu^{\prime 3}$ are stable.\footnote{The parities allow for a decay of one DM component into two others, however this is forbidden kinematically. }

\paragraph{Gauge boson - scalar mixing}

According to Eq.~(\ref{removemix}), the $A_\mu \partial^\mu \varphi$ mixing terms are removed by the redefinition (which does not affect the gauge boson masses)
\bal
\tilde A^6_\mu&=A^6_\mu+ \partial_\mu Y^6 \;, \quad \textrm{where}\quad Y^6=\frac2{\tilde g} \frac{v_2}{v_1^2+v_2^2} \varphi^4\;,
\nn
\\
\tilde A^7_\mu&=A^7_\mu+ \partial_\mu Y^7 \;, \quad \textrm{where}\quad Y^7=-\frac2{\tilde g} \frac{v_2}{v_1^2+v_2^2} \varphi^3\,,
\eal
while, according to Eq.~(\ref{normalizescalars}), the canonically normalised scalars are
\be
\tilde \varphi^3 = \frac{v_1}{\sqrt{v_1^2+v_2^2}} \varphi^3 \,,
\quad
\tilde \varphi^4 = \frac{v_1}{\sqrt{v_1^2+v_2^2}} \varphi^4 \,.
\ee

\paragraph{Scalar masses}
The general scalar potential has many parameters. To make our discussion more transparent, let us assume the symmetry 
 $\phi_2 \to - \phi_2$ which does not affect the essence of our considerations and 
 requires $m_{12}^2=\lambda_{H12}=\lambda_6=\lambda_7=0$.
In this case, it turns out that the potential has no local minima with all of $v,v_1,v_2, v_3$ being nonzero such that setting $v_3=0$ is actually required.
It should be noted that in practice we are considering the limit in which 
the above quantities are very small but nonzero such that the decay channels for heavy particles, e.g. $\tilde\varphi_3$, into the SM fields are open.

The VEVs $v,v_1,v_2$ can be expressed in a compact form using the matrix
\bal
\mathbf \Lambda \equiv
\left( \begin{array}{ccc} 
\lambda_H & \lambda_{H11} & \lambda_{H22} \\
\lambda_{H11} & \lambda_1 & \lambda_3 \\
\lambda_{H22} & \lambda_3 & \lambda_2 
\end{array} \right) 
\eal
as well as the matrices $\mathbf \Lambda_{ij}$ defined as $(-1)^{i+j}$ times the matrix obtained by deleting the $i$-th row and $j$-th column of $\mathbf \Lambda$, i.e.~$\det \mathbf \Lambda_{ij}$ is the $(i,j)$-cofactor of $\mathbf \Lambda$.
One finds
\bal
v^2&= - 2 ( m_H^2 \det{\mathbf \Lambda_{11}} +m_{11}^2 \det{\mathbf \Lambda_{21}}+m_{22}^2 \det{\mathbf \Lambda_{31}})/\det{\mathbf \Lambda} \,, \nn
\\
v_1^2&= -2 ( m_H^2 \det{\mathbf \Lambda_{12}} +m_{11}^2 \det{\mathbf \Lambda_{22}}+m_{22}^2 \det{\mathbf \Lambda_{32}})/\det{\mathbf \Lambda} \,, \nn
\\
v_2^2&= -2 ( m_H^2 \det{\mathbf \Lambda_{13}} +m_{11}^2 \det{\mathbf \Lambda_{23}}+m_{22}^2 \det{\mathbf \Lambda_{33}})/\det{\mathbf \Lambda} \,.
\eal
The mass terms for the scalars are
\be
- \Lcal \supset \frac12 \Phi^T \mathbf{m}^2 \Phi + \frac14 (\lambda_4+\lambda_5) (v_1^2+v_2^2) \, \tilde \varphi_3^2 + \frac14 (\lambda_4-\lambda_5) (v_1^2+v_2^2) \, \tilde \varphi_4^2 \;,
\label{phi4-mass}
\ee
where $\Phi^T\equiv (h,\varphi_1,\varphi_2)$ and
\bal
\mathbf{m}^2 =
\left( \begin{array}{ccc} 
\lambda_H v^2 & \lambda_{H11} v v_1 & \lambda_{H22} v v_2 \\
\lambda_{H11} v v_1 & \lambda_1 v_1^2& \lambda_3 v_1 v_2\\
\lambda_{H22} v v_2& \lambda_3 v_1 v_2 & \lambda_2 v_2^2\\
\end{array} \right) \,.
\eal
Note that $\tilde \varphi_3$ and $\tilde \varphi_4$ are generally heavier than $A^1,A^2,A'^3$ since their masses involve $v_1$
(unless $\lambda_{4,5}$ are very small). 

The matrix $\mathbf{m}^2$ is positive definite if and only if $\det \mathbf \Lambda>0$, $\det \mathbf \Lambda_{33}\equiv \lambda_H \lambda_1- \lambda_{H11}^2>0$ and $\lambda_H>0$.
It can be diagonalised by an orthogonal transformation
$O^T \mathbf{m}^2 O = \textrm{diag}(m_{H_1}^2,m_{H_2}^2,m_{H_3}^2) ,$
where
\bal
O&= \left( \begin{array}{ccc}
c_{12} c_{13}& s_{12} & c_{12}s_{13}\\
-c_{13} c_{23} s_{12} - s_{13} s_{23} & c_{12} c_{23} &-c_{23} s_{12} s_{13}+ c_{13} s_{23} \\
-c_{23} s_{13} + c_{13} s_{12} s_{23} &-c_{12} s_{23} &c_{13} c_{23} + s_{12} s_{13} s_{23}
\end{array} \right) \;
\eal
and we have used the abbreviation $s_{ij} \equiv \sin \theta_{ij}, c_{ij} \equiv \cos \theta_{ij}$. Instead of providing the most general formulae, let us focus on a simplified case.
 Suppose that the $(12)$ and $(23)$ entries of $\mathbf{m}^2 $ are much smaller than the other matrix elements, in other words, that $\lambda_{H11}$
and $\lambda_3$ are very small. In this case, the Higgs mixing with
$\varphi_2$ is the dominant one. The reason behind this choice is that 
the DM constituents $A^1,A^2$ and $A'^3$ all have a significant coupling to $\varphi_2$, see Table~\ref{AAscalar}.
Given that the mixing between the Higgs and $\varphi_2$ is substantial,
this facilitates DM annihilation into the SM fields. 
Clearly, similar considerations also apply to the general case with all the angles $\theta_{ij}$ being significant.

Assuming small $\mathbf{m}^2_{12},\mathbf{m}^2_{23}$, one finds
\besub
\bal
& \theta_{12} \approx p_{32}\, \mathbf{m}^2_{12} + q \, \mathbf{m}^2_{23} \;,
\\
& \theta_{23} \approx p_{21}\, \mathbf{m}^2_{23} + q \, \mathbf{m}^2_{12} \;,
\\
& \tan 2 \theta_{13} \approx \frac{2 \lambda_{H22} v v_2}{\lambda_2 v_2^2-\lambda_H v^2} \,,
\eal
\eesub
where $p_{ij}=(\mathbf{m}^2_{ii}-\mathbf{m}^2_{jj})/s$, $q=\mathbf{m}^2_{13}/s$ and $s=(\mathbf{m}^2_{13})^2+(\mathbf{m}^2_{11}-\mathbf{m}^2_{22})(\mathbf{m}^2_{22}-\mathbf{m}^2_{33})$.
The mass eigenstates are
\bal
\left( \begin{array}{c}
h_1\\
{\cal H}\\
h_2
\end{array} \right)
\equiv O^T \Phi
\approx
 \left( \begin{array}{c}
c_{13} h - s_{13} \varphi_2 \\
\varphi_1 \\
c_{13} \varphi_2 + s_{13} h
\end{array} \right)
- \theta_{12}
 \left( \begin{array}{c}
 c_{13} \varphi_1 \\
-h \\
s_{13} \varphi_1
\end{array} \right)
- \theta_{23}
 \left( \begin{array}{c}
 s_{13} \varphi_1 \\
 \varphi_2\\
-c_{13} \varphi_1
\end{array} \right)
\,. \label{rotation-matrix}
\eal
and the mass--squared eigenvalues are 
\besub
\bal
& m^2_{h_1,h_2} \approx \frac12 ( \lambda_2 v_2^2+\lambda_H v^2) \mp \frac{ \lambda_2 v_2^2-\lambda_H v^2}{2 \cos 2 \theta_{13}}
\\
& m^2_{\cal H} \approx \lambda_1 v_1^2 \,.
\eal
\eesub
The eigenstates $h_1,h_2$ are typically the lighter ones, while ${\cal H}$
is heavy (unless $\lambda_1$ is very small). In our analysis of DM phenomenology, we retain only the former states.
In summary, the relevant light fields are the DM components $A^1,A^2$ and $A'^3$ as well as the mediators $h_1,h_2$ which link the dark sector to the SM fermions and gauge bosons.

\subsubsection{Couplings}

The full list of the couplings is not necessary for our DM studies.
The important couplings are those with two gauge bosons and one or two scalars at the vertex. In terms of the variables $\tilde A_\mu^a $
and $\tilde \varphi^i$, most of these are listed in Table~\ref{AAscalar} 
and Table~\ref{AAscalarscalar}. For our applications, the couplings of $h_1$ and $h_2$ are obtained from these tables using the relation
 (\ref{rotation-matrix}), in which one may neglect $\theta_{12}$
 and $\theta_{23}$.

We focus on the case $v_1 \gg v_2$ so that
 DM consists of $A^1,A^2$ and $A'^3$. Other fields with non--trivial parities
 decay into these states and the SM particles. For instance, 
 the processes $\tilde \varphi_4 \rightarrow A^1 A^2 \, +$~SM and 
 $\tilde \varphi_4 \rightarrow A^{3\prime}\, +$~SM are allowed. 
 (When $v_1$ and $v_2$ are close, the composition of DM depends on the mass splittings.)
 DM annihilation and scattering proceeds through an exchange of $h_1$
 and $h_2$. Therefore, only the vertices involving these fields 
 play a significant role.

%=========================================================================
%=========================================================================
\section{Generalisation to arbitrary SU(N)} \label{SUN}
%=========================================================================
%=========================================================================
SU(N) is broken completely by generic VEVs of $N-1$ fields $\phi_i$ in the fundamental representation.
The $\phi_i$'s can be gauge--transformed to the form
\begin{equation}
\phi_1= \left( \begin{matrix} 
   0\\ 
   0 \\
   ...\\
   0 \\
   \rho_1 
  \end{matrix}
\right) ~~,~~
\phi_2= \left( \begin{matrix} 
   0\\ 
   0 \\
   ...\\
   \rho_2^{(1)} \\
   \rho_2^{(2)} e^{i \xi_2}
   \end{matrix} 
\right) ~~,~~...~~,~~
\phi_{N-1}= \left( \begin{matrix}
    0\\ 
   \rho_{N-1}^{(1)} \\
   ...\\
   \rho_{N-1}^{(N-2)} e^{i \xi_{N-1}^{(N-3)}}\\ 
   \rho_{N-1}^{(N-1)} e^{i \xi_{N-1}^{(N-2)}}
   \end{matrix} 
\right) ~~.
\end{equation}
Here the radial fields $\rho_i^{(j)}$ and the phases $\xi_i^{(j)}$ are real. We label the scalars such that 
the lightest gauge fields are associated with the SU(2) subgroup which gets broken
at the last stage by a VEV of $\rho_{N-1}^{(1)}$. We assume that the VEVs as well as the couplings in the scalar potential are all real so that {\it CP} is preserved in the hidden sector.

The generalisation of the $Z_2 \times Z_2'$ parity to SU(N) is as follows. 
The transformation properties of the gauge fields are identified with those of the corresponding SU(N) generators. 
The basis of the $N(N-1)$ off--diagonal generators $T^{ab}, \tilde T^{ab}$ can be chosen as 
\begin{eqnarray}
&& ( T^{ab} )_{ij}= \delta_{ia} \delta_{jb} + \delta_{ib} \delta_{ja} \;, \nonumber\\
&& ( \tilde T^{ab})_{ij}= -i \delta_{ia} \delta_{jb} + i \delta_{ib} \delta_{ja} \;,
\end{eqnarray}
where $a=1,..,N-1$ and $b=2,..,N$.
With the Cartan generators denoted by $H^\alpha$, the $Z_2$ associated with 
complex conjugation of the group elements acts as
\begin{eqnarray}
&& T^{ab} \rightarrow -T^{ab} ~~,~~ \tilde T^{ab} \rightarrow \tilde T^{ab}
~~,~~ H^\alpha \rightarrow - H^\alpha \;. \label{Z21N}
\end{eqnarray}
This is a well known outer automorphism of SU(N) which entails the corresponding symmetry of the Yang--Mills Lagrangian.

Another $Z_2$ can be defined by reflecting the off--diagonal generators containing nonzero elements in the first row:
\begin{eqnarray}
&& T^{1a} \rightarrow -T^{1a} ~~,~~ \tilde T^{1a} \rightarrow - \tilde T^{1a}
~~, \nonumber\\
&& T^{bc} \rightarrow T^{bc} ~~~~~,~~ \tilde T^{bc} \rightarrow \tilde T^{bc}
~~ ~~~(b,c \geq 2), \nonumber\\
&& H^\alpha \rightarrow H^\alpha \;. \label{Z22N}
\end{eqnarray}
It is easy to show that this $Z_2$ is an inner automorphism. Indeed, it corresponds to the group transformation with
\begin{equation}
U= e^{ { i \pi \over N}} \; {\rm diag}(-1,1,...1) \;. \label{U}
\end{equation}
The gauge fields $A^{1-3}_\mu$ associated with the 
upper left SU(2) block 
\begin{equation}
T^{12},~~\tilde T^{12},~~ H^1= {\rm diag}(1,-1,0,...,0)
\end{equation}
transform under these parities the same way as did the SU(2) gauge fields
under $Z_2 \times Z_2'$ of the previous section. The above gauge transformation of course leaves the Yang--Mills Lagrangian invariant.

These symmetries are preserved by gauge interactions with scalars in our set--up. The $Z_2$
associated with complex conjugation acts on scalars by reflecting the complex phases,
which therefore correspond to odd fields under the transformation~(\ref{Z21N}). This symmetry is guaranteed by {\it CP} invariance of the Lagrangian and preserved by the vacuum (assuming no spontaneous {\it CP} violation).
The second $Z_2$ is a gauge transformation. On vectors of the form 
$(0,a_1,...,a_{N-1})$, it acts as multiplication by an overall 
constant phase which cancels in all the Lagrangian terms. It is therefore a valid symmetry in the broken phase as well. 

As long as the $\phi_i$ have a zero first component,
the interaction vertices contain an even number of $T^{1a}$ and $\tilde T^{1a}$.
 The gauge fields associated with $a>2$ are heavier than those corresponding to $a=2$. By virtue of 
 the vertices involving $T^{12} T^{1k}$ $(k>2)$ and the matter fields, they decay 
 to the lighter fields such that only the final SU(2) block remains stable. 
 (Similar considerations also apply to other heavy gauge fields.)
Then DM is composed mostly of the aforementioned $A^{1-3}_\mu$ whose stability is enforced by $Z_2 \times Z_2'$.\footnote{As before, we take the ``phase'' fields to be heavy since they get
their masses from large VEVs $\langle \phi_i \rangle$, unlike $A^{1-3}_\mu$. In the presence of more than $N-1$ fundamentals, this logic no longer applies and the light phase fields can constitute DM.} 
 
As in the SU(3) case, DM consists of three components, two of which are degenerate in mass,
\begin{equation}
m_{ A^{1\prime }} =m_{ A^{2\prime }} \not= m_{ A^{3\prime }} \;,
\end{equation}
where $ A^{1\prime -3 \prime}_\mu$ are the mass eigenstates consisting mostly of $ A^{1-3 }_\mu$
with some admixture of other gauge fields (see the SU(3) example). The degeneracy persists at loop level. This can be seen as follows. The SU(N) Lie algebra possesses a discrete symmetry which interchanges the real and imaginary generators with nonzero entries in the first row,
\begin{equation}
T^{1a} \rightarrow \tilde T^{1a} ~~,~~ \tilde T^{1a} \rightarrow - T^{1a}~~,
\label{A1-A2}
\end{equation}
while all the other generators remain intact.
This is achieved by the group transformation
\begin{equation}
U'= e^{ { -i \pi \over 2 N}} \; {\rm diag}(i,1,...1) \;,
\end{equation}
which can be recognised as the square root of $U$ in (\ref{U}).
This gauge transformation acts on $\phi_i$ with a zero first entry as an overall
constant phase multiplication. Since such a phase cancels in all of the Lagrangian terms,
(\ref{A1-A2}) remains a valid symmetry even in the broken phase.

 Consider now the mass matrix for the gauge fields associated with $T^{1a}$ and $\tilde T^{1a}$. By virtue of (\ref{Z22N}), only fields corresponding to $T^{1a}$ and $\tilde T^{1a}$
 can mix, while (\ref{Z21N}) forbids a mixing between the tilded and untilded fields.
 The resulting mass matrix for $T^{1a}$ is then identical to that of $\tilde T^{1a}$
 according to (\ref{A1-A2}). Analogous considerations apply to the kinetic terms.
 Hence the lightest eigenstates have the same mass. 
 
 The resulting DM phenomenology is analogous to that for the SU(3) case.

%=========================================================================
%=========================================================================
\section{Dark matter phenomenology }
%=========================================================================
%=========================================================================

 In what follows, we consider direct and indirect detection as well as relic abundance constraints on vector DM. 
 In the U(1) and SU(2) cases, all the scattering processes are mediated by $h_1$ and $h_2$. For SU(3) and larger groups, further states can contribute.
However, we make the simplifying assumption that the Higgs mixing with those states is small and/or such states are heavy. In that case, it suffices to consider an exchange of $h_1$ and $h_2$ only. We note that earlier phenomenological analyses of vector dark matter 
have appeared in various contexts~\cite{Kanemura:2010sh}-\cite{Duch:2014xda}.
The current collider searches for dark matter, e.g. in the form of a monojet plus missing energy, do not set further useful constraints on the model, see e.g. Ref.~\cite{Kim:2015hda}.

%=========================================================================
\subsection{U(1) dark matter}
%=========================================================================
Let us start with the U(1) case. The relevant terms in the Lagrangian are
\bal
{\cal L} &\supset {1\over 2} m_A^2 A_\mu A^\mu + {\tilde g \, m_A \over 2} 
\left(-h_1 \sin\theta + h_2 \cos\theta \right) A_\mu A^\mu \nonumber \\
&+
{\tilde g^2 \over 8 } \left( h_1^2 \sin^2 \theta -2 h_1 h_2 \sin\theta \;\cos\theta +
h_2^2 \cos^2 \theta \right) A_\mu A^\mu \;.
\label{DMinteractions}
\eal
The couplings of $h_1$ and $h_2$ to the SM fields are those of the SM Higgs up
to the suppression factors of $\cos\theta$ and $\sin\theta$, respectively.
A phenomenological analysis of this model in the decoupling limit $m_{h_2} \gg m_{h_1}$,
$\sin\theta \rightarrow 0$ can be found in~\cite{Kanemura:2010sh} and 
\cite{Lebedev:2011iq,Djouadi:2011aa}. 
Related studies have also appeared in~\cite{Farzan:2012hh,Baek:2012se}.

The DM scattering on nucleons proceeds through the $t$--channel exchange of $h_{1,2}$
 and leads to the following spin--independent cross
section (see e.g.~\cite{Lebedev:2011iq}),
\begin{equation}
\sigma^{\rm SI}_{A-N}= {g^2 \tilde g^2 \over 16 \pi} \; {m_N^4 f_N^2 \over m_W^2} \;
{ (m_{h_2}^2 - m_{h_1}^2 )^2 \sin^2 \theta \; \cos^2 \theta \over m_{h_1}^4 m_{h_2}^4} \;,
\end{equation}
where $m_N$ is the nucleon mass and $f_N \simeq 0.3$ parametrizes the Higgs--nucleon coupling. One should keep in mind that there is significant uncertainty in $f_N$ and here
we use the value somewhat smaller than the one assumed in~\cite{Lebedev:2011iq}. 
As expected, $h_1$ and $h_2$ contribute with opposite signs. Since 
$m_{h_2}^2 \gg m_{h_1}^2$ in realistic cases, the cancellation is not very significant.
\begin{figure}[h] 
\centering{
\includegraphics[scale=0.283]{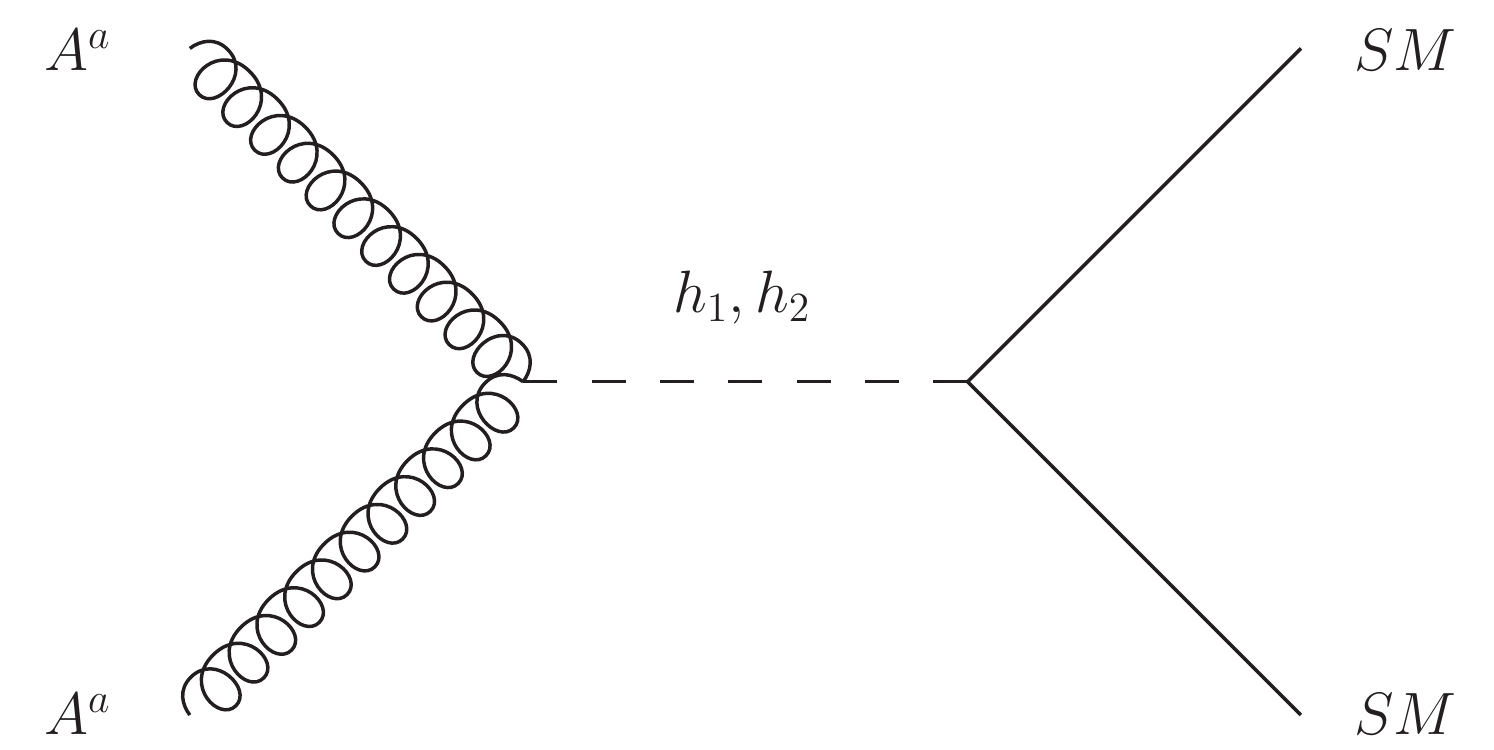}
\includegraphics[scale=0.283]{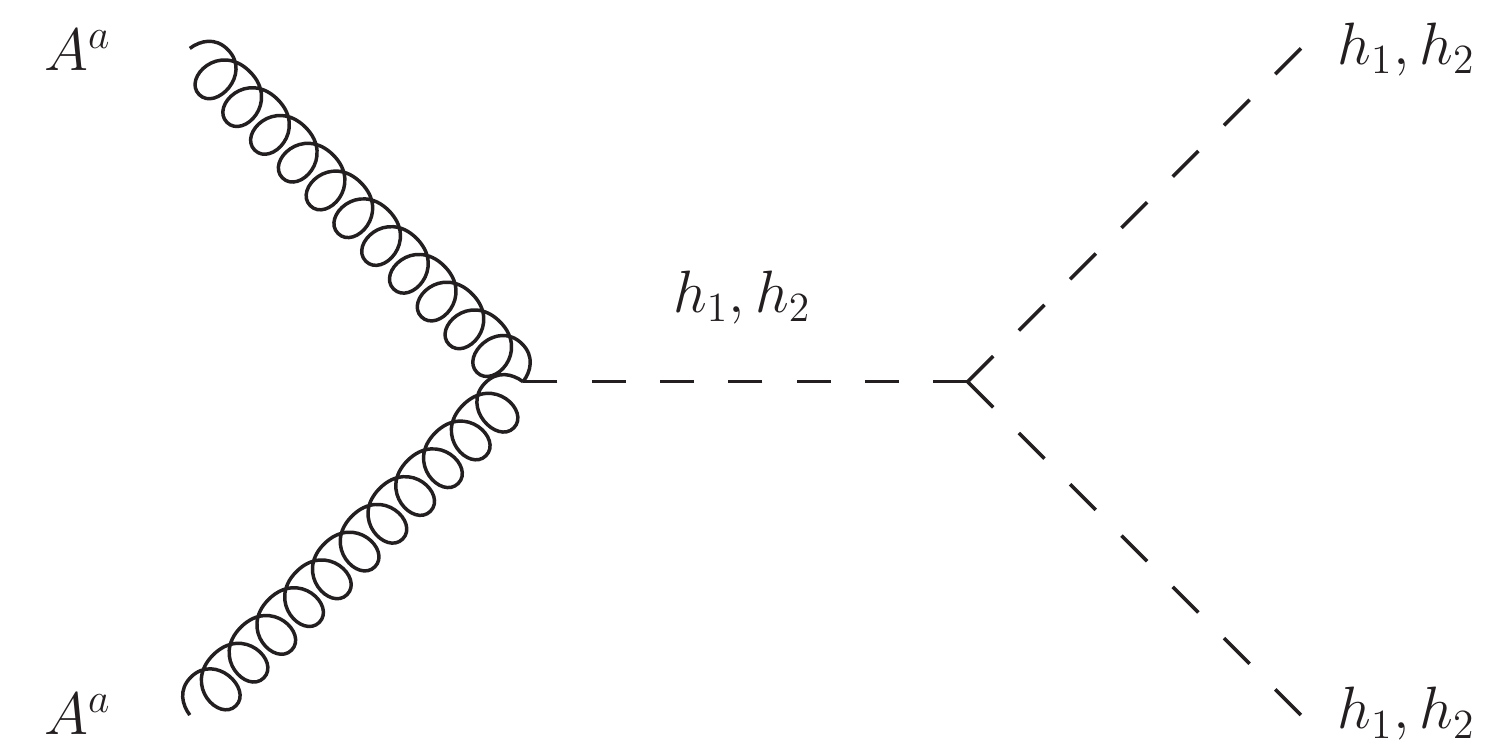}
\includegraphics[scale=0.283]{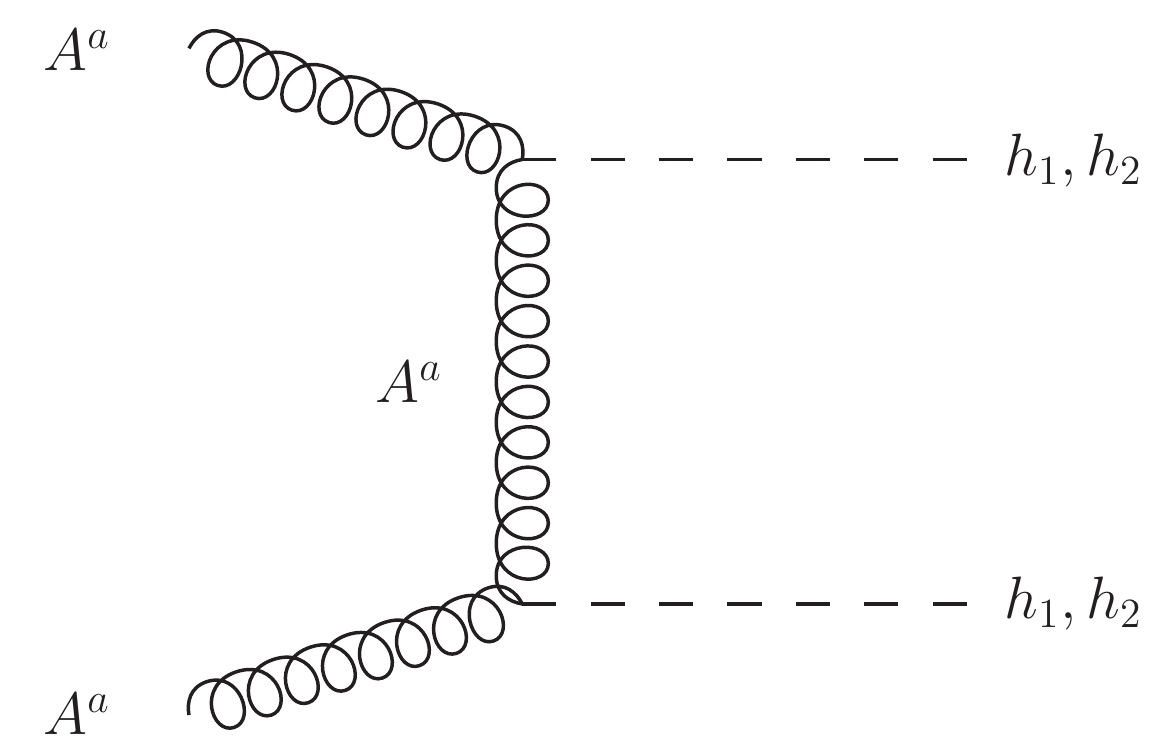}
\includegraphics[scale=0.283]{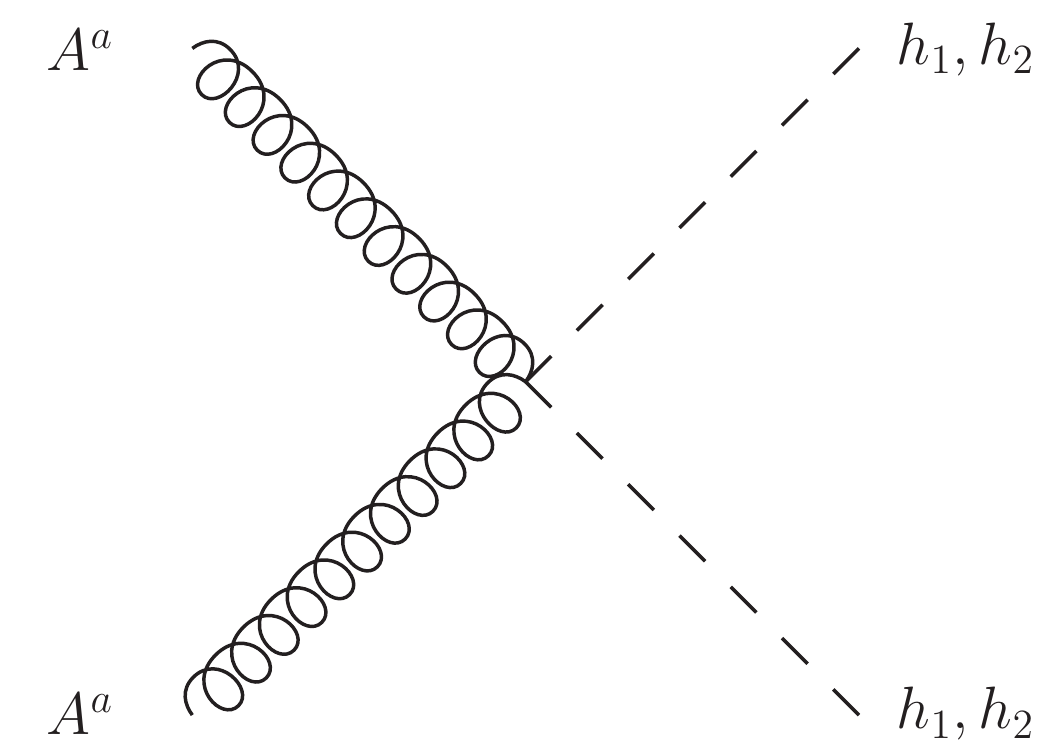}
 }
\caption{ \label{diagrams1}
Leading diagrams for vector DM annihilation.
}
\end{figure}

The calculation of the DM annihilation cross section is more involved due to a few contributing diagrams (Fig.~\ref{diagrams1}). To compute the DM relic abundance, we use the software package micrOMEGAs 4.1.8~\cite{Belanger:2014vza}. It is important to note that the leading $s$--channel annihilation amplitude
due to the $h_{1,2}$--exchange is proportional to the factor
 \begin{equation}
 {\cal A}^{\rm annih} \propto \sin\theta \;\cos\theta \; \left( {1\over s-m_{h_1}^2 } -
 {1\over s-m_{h_2}^2 } \right) \;,
\end{equation}
where the $s$--parameter can be approximated by $s \simeq 4 m_A^2$. Therefore,
the amplitude is highly suppressed at $m_A \gg m_{h_2}$. 
Although other annihilation channels remain available, this makes DM annihilation
inefficient for heavy masses and the corresponding parameter space is challenged by the direct detection constraint. 

Clearly, the $s$--channel annihilation becomes very efficient around the resonances,
$m_A \simeq m_{h_1}/2$ and $m_A \simeq m_{h_2}/2$. In this case, a very small gauge coupling is sufficient to obtain the right relic abundance. 
The first resonance is quite narrow due to the small width of the SM Higgs, whereas 
the second resonance is broad since many decay channels are available to $h_2$. For 
$m_{h_2} > 2 m_{h_1}$, the decay $h_2 \rightarrow h_1 h_1$ becomes important. Its significance depends on $\lambda_{h \phi}$ of Eq.~(\ref{eq1}) with BR${_{h_2 \rightarrow h_1 h_1}}$ increasing for larger $\lambda_{h \phi}$ (see the explicit formulae in~\cite{Falkowski:2015iwa}). The resonance is widened further by the thermal averaging over the DM momentum.

The most important features of DM annihilation are associated with the $s$--channel.
Other channels play a less significant role. Similar considerations apply to the indirect detection constraint due to the gamma ray emission in the process of DM annihilation (FERMI). 

The last constraint we impose is of theoretical nature. We require that the theory be perturbative at the TeV scale. One way to enforce it is to demand perturbative unitarity at tree level, for instance, in the process $h_i h_i \rightarrow h_j h_j$. The resulting constraint was estimated in~\cite{Chen:2014ask} to be 
\begin{equation}
\lambda_i < {\cal O} (4\pi/3) \;,
\label{unitarity}
\end{equation} 
where $\lambda_i$ are the scalar quartic couplings.
We define the quartic couplings involving $\phi$ by 
$\Delta V_{\rm quart} = \lambda_{h \phi} \vert H \vert^2 \vert \phi \vert^2 +
{1\over 2} \lambda_{\phi} \vert \phi \vert^4 $. They can be expressed in terms of 
the masses, the gauge coupling and the mixing angle as (see e.g.~\cite{Falkowski:2015iwa}) 
\begin{eqnarray}
&& \lambda_{h \phi}= \tilde g ~\sin 2 \theta~ { m_{h_2}^2-m_{h_1}^2 \over 4 v m_A } \;,
\nonumber \\ 
&& \lambda_{\phi}= \frac{ 4 \, \lambda_{h \phi}^2 }{\sin^2 2 \theta } \frac{v^2}{m_{h_2}^2-m_{h_1}^2} \left(\frac{m_{h_2}^2}{m_{h_2}^2-m_{h_1}^2} -\sin^2 \theta \right) \;,
\end{eqnarray} 
 where we have used $\tilde v =2 m_A/ \tilde g $. 
This implies that both $\lambda_{h \phi}$ and $\lambda_\phi$ become large for heavy $h_2$ 
or for light dark matter. As a result, Eq.~(\ref{unitarity}) imposes an important constraint 
on our model. We further require the standard perturbativity constraint 
\begin{equation}
{\tilde g^2 \over 4 \pi} < 1 ~,
\end{equation} 
which we find less significant for our purposes. 
\begin{figure}[h] 
\centering{
\includegraphics[scale=0.43]{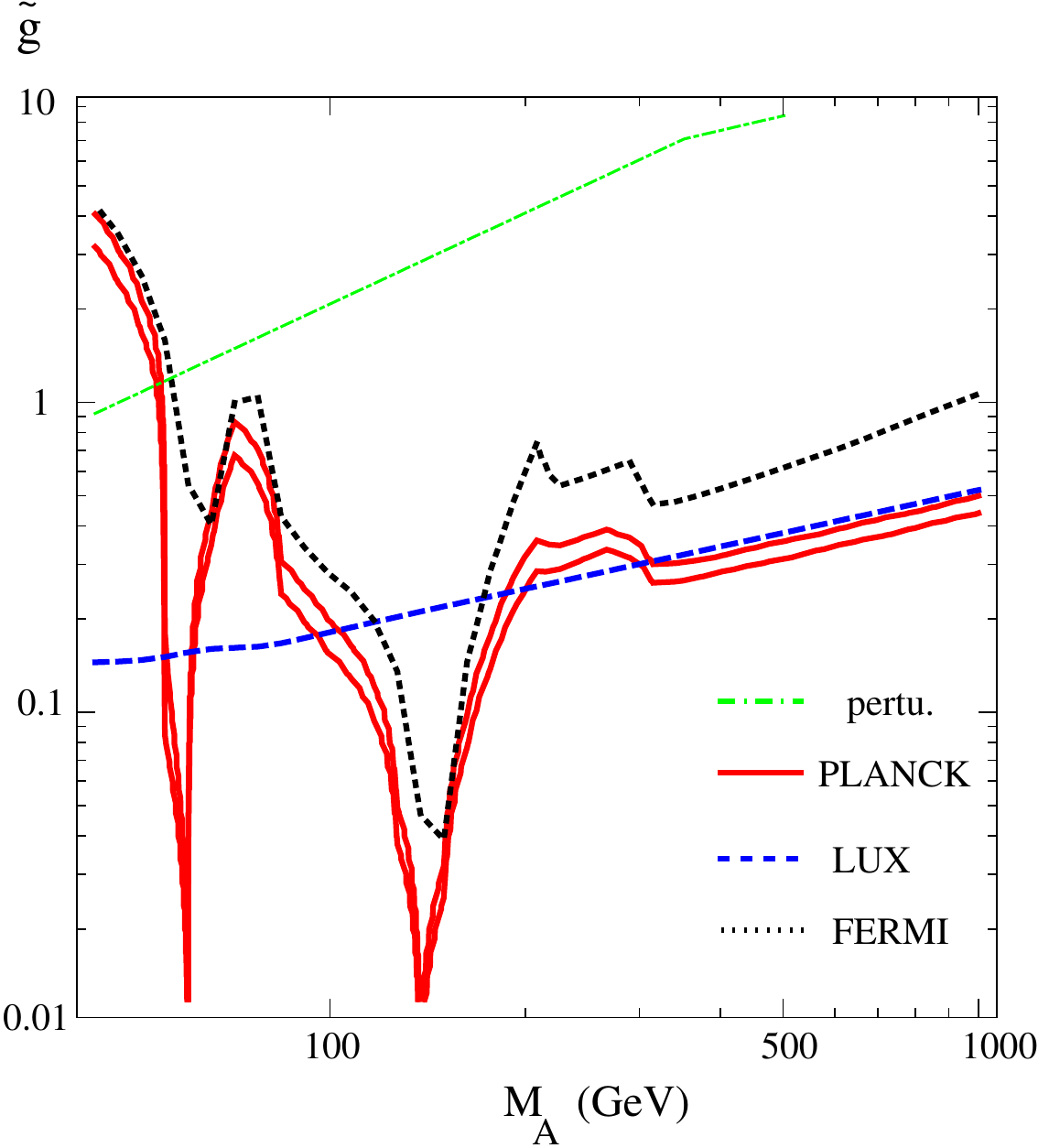}
\includegraphics[scale=0.43]{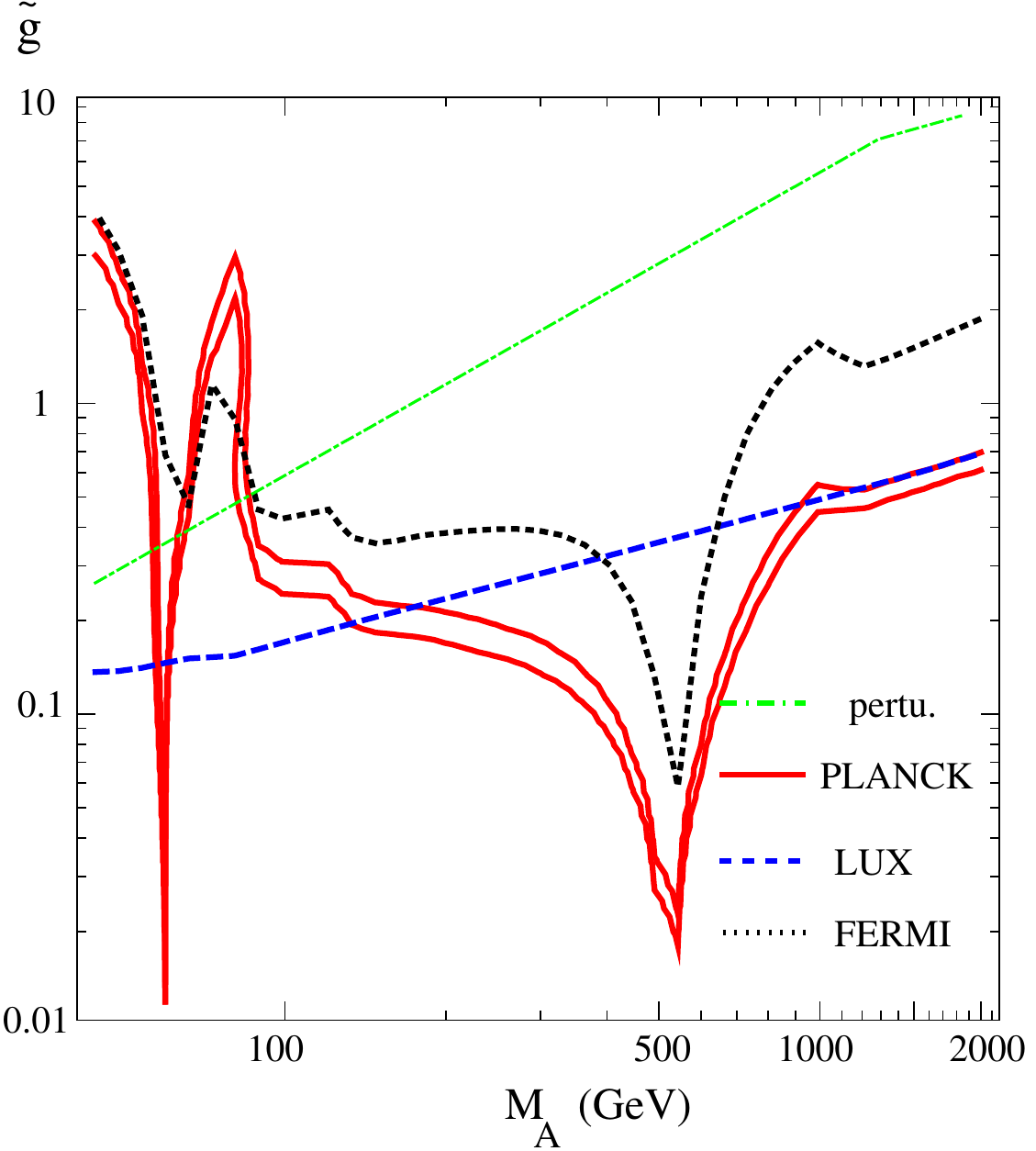}
\includegraphics[scale=0.43]{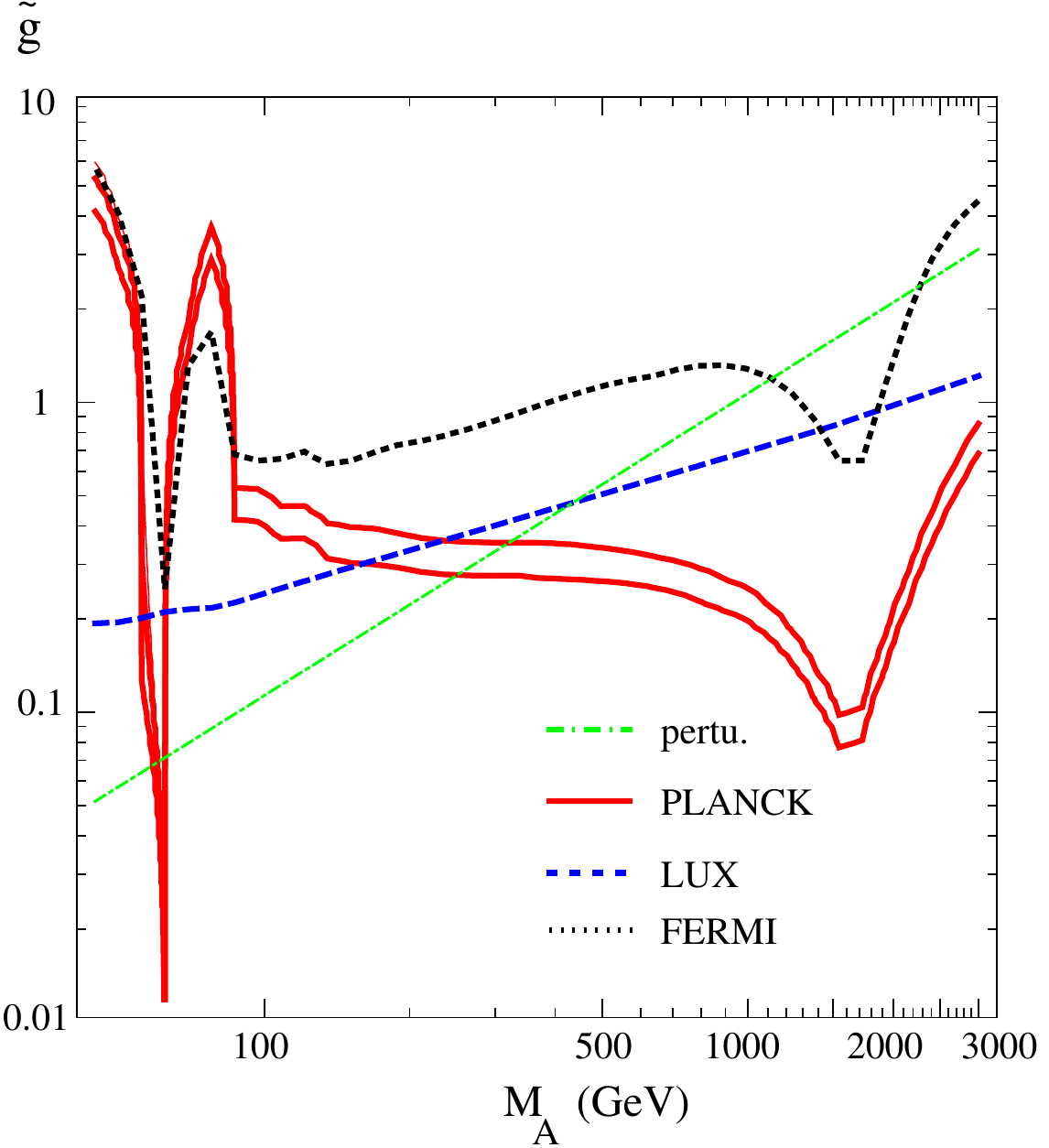}
\includegraphics[scale=0.43]{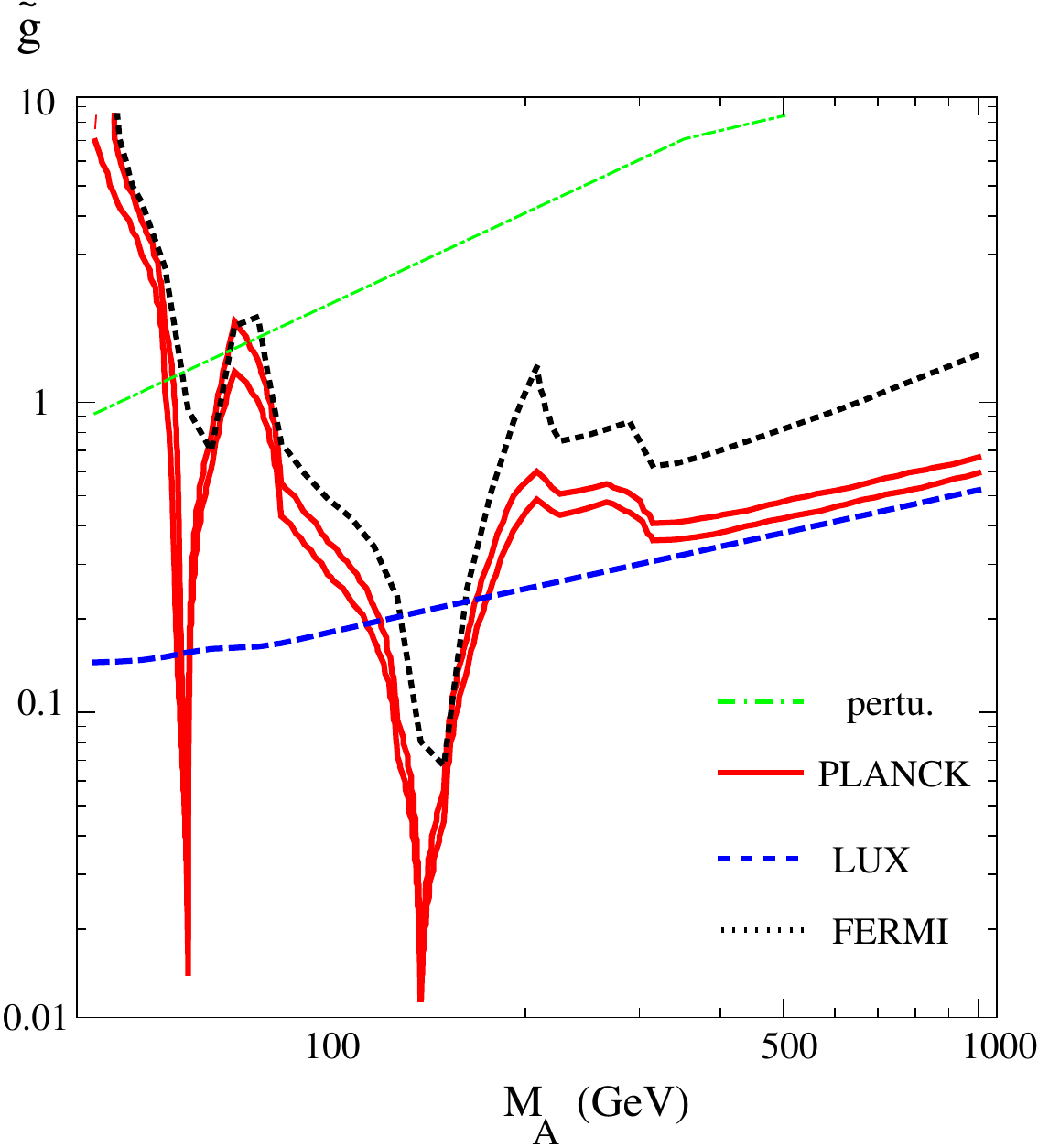}
\includegraphics[scale=0.43]{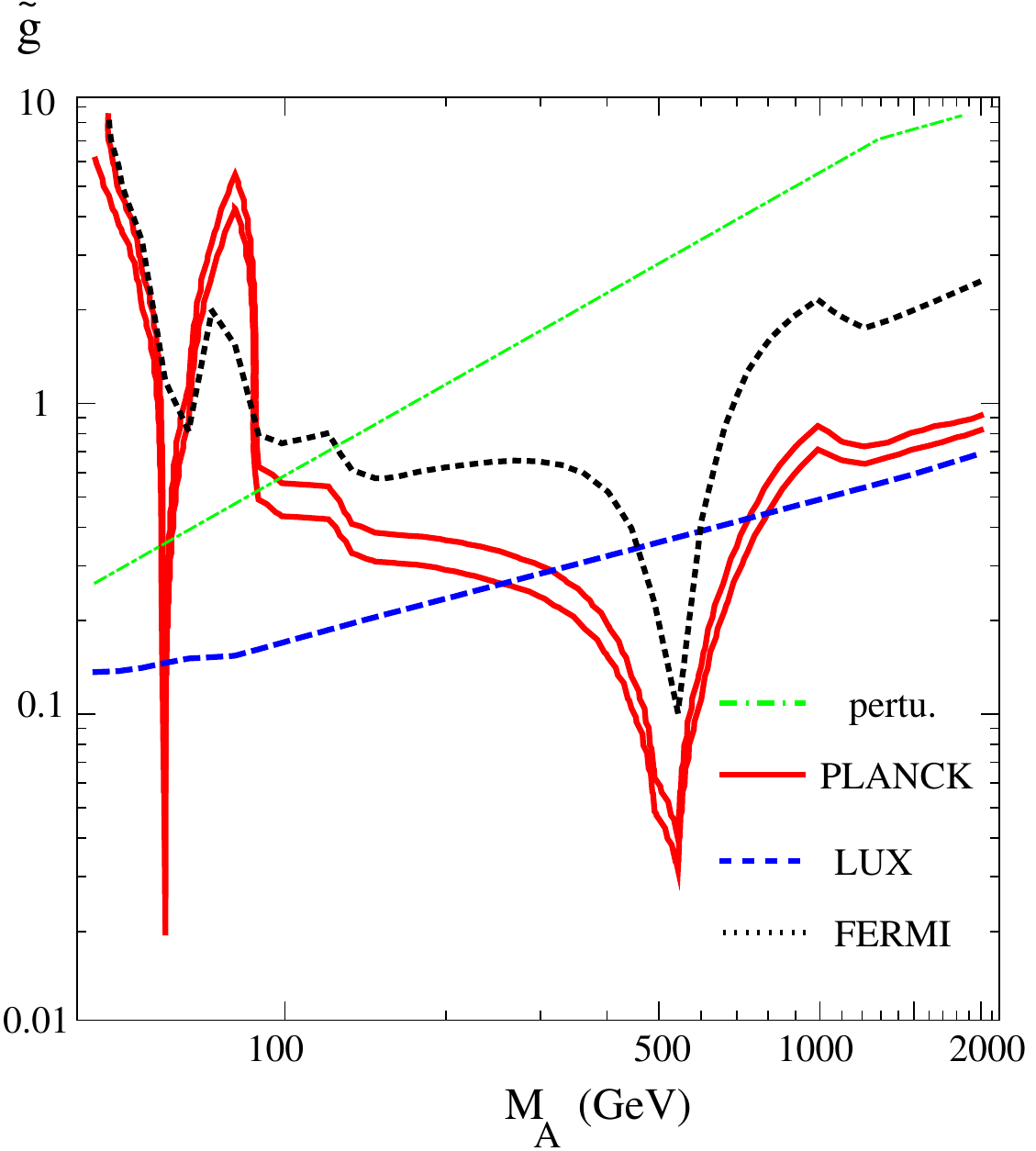}
\includegraphics[scale=0.43]{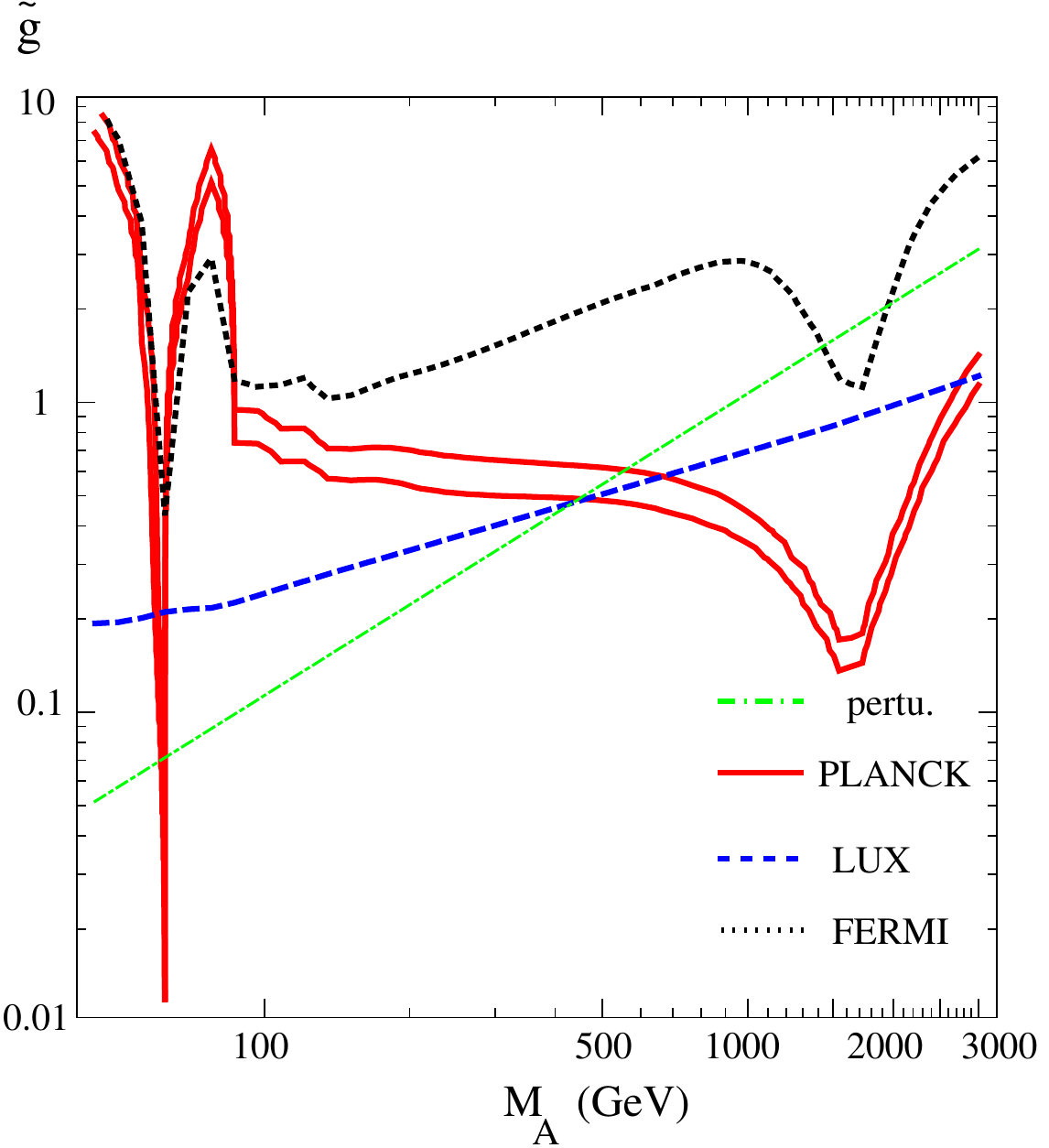}
}
\caption{ \label{plots}
Constraints on the gauge coupling $\tilde g$ vs DM mass $m_A$ for U(1) and SU(2).
The area between the red lines is favoured by the DM relic abundance, while the regions above 
the dashed blue, dotted black and green lines are ruled out by direct detection, indirect detection and perturbativity of $\lambda_i$, respectively.
{\it Upper row:} U(1) dark matter with $\sin\theta=0.3, m_{h_2}= 280$ GeV (left),
 $\sin\theta=0.3, m_{h_2}= 1$ TeV (center), $\sin\theta=0.2, m_{h_2}= 3$ TeV (right).
 {\it Lower row:} SU(2) dark matter with $\sin\theta=0.3, m_{h_2}= 280$ GeV (left),
 $\sin\theta=0.3, m_{h_2}= 1$ TeV (center), $\sin\theta=0.2, m_{h_2}= 3$ TeV (right). 
}
\end{figure}

Our results for U(1) DM are presented in Fig.~\ref{plots}, upper row. 
We include the constraints from PLANCK~\cite{Ade:2015xua} (relic abundance), LUX~\cite{Akerib:2013tjd} (direct detection),
FERMI~\cite{Ackermann:2015zua} (indirect detection) and perturbativity of $\lambda_i$.
The area between the red lines is consistent with the thermal relic DM abundance measured by PLANCK. The LUX data provide the strongest constraint on the allowed parameter space. 
The FERMI bound is typically relevant for light DM, while the perturbativity bound becomes
important for heavy $h_2$. 
In each panel, the mixing angle is chosen such that, for a given $m_{h_2}$,
it is consistent with the LHC and EW precision data~\cite{Falkowski:2015iwa}. Heavier
$h_2$ imply smaller $\sin\theta$, so we take $\sin\theta=0.3$ for the left and center panels,
and $\sin\theta=0.2$ for the right panel.

The two dips are associated with the resonant annihilation through $h_1$ and $h_2$. The second
resonance gets broader with increasing $h_2$ due to the increase in $\lambda_{h \phi}$ and
availability of the decay $h_2 \rightarrow h_1 h_1$. The area around this resonance constitutes the largest parameter space consistent with all of the constraints. For $m_{h_2}= 280$ GeV,
the allowed DM mass range is about 100 GeV; for $m_{h_2}= 1$ TeV, it widens to 1 TeV, and
for $m_{h_2}= 3$ TeV, it reaches more than 3 TeV. The resonance is broadened by the thermal averaging
over the DM momentum, so even though it appears very broad for $m_{h_2}= 3$ TeV,
it is still consistent with perturbativity. 

The dip associated with the resonant annihilation through $h_1$ is quite narrow and does not open up further significant areas of parameter space. 
Other features of the PLANCK curve are local peaks corresponding to the kinematic opening of additional annihilation channels. For instance, the peak at $m_A \sim 80$ GeV is associated with the $W^+ W^-$ final state. There are further visible peaks at $(m_{h_2} + m_{h_1})/2$ and $m_{h_2}$.

Away from the resonances, there appears to be a further allowed region at $m_A > m_{h_2}$.
Since the $h_2$ production is not suppressed by $\sin\theta$, 
in this case the $t$--channel annihilation $AA \rightarrow h_2 h_2$
and the quartic interactions dominate.
The Planck--allowed strip is dangerously close to the LUX bound, so the conclusion
depends strongly on the nucleon--Higgs coupling $f_N \simeq 0.3$, which suffers from substantial uncertainties.

%=========================================================================
\subsection{SU(2) dark matter}
%=========================================================================
Aside from the gauge self--interactions, the Lagrangian (\ref{DMinteractions}) applies
to the SU(2) case as well, up to the summation over the 3 species, $A_\mu A^\mu 
\rightarrow A_\mu^a A^{a \mu }$. The main change compared to the U(1) case is that 
the annihilation cross section decreases since only the species with the same group index 
can annihilate through the Higgs--like states. In order to keep the same relic abundance, one needs to increase the gauge coupling. Since the $s$--channel annihilation through $h_{1,2}$
often dominates, this amounts approximately to
\begin{equation}
\tilde g \rightarrow \sqrt{3} \tilde g \;,
\end{equation}
whereas if 
 the $t$--channel and the quartic interactions dominate, the rescaling factor is closer
 to $3^{1/4}$. The direct detection constraint remains the same as in the U(1) case since particles with different group indices scatter the same way on nucleons.
This decreases somewhat the allowed parameter space compared to the Abelian case (Fig.~\ref{plots}).

\begin{figure}[h] 
\centering{
\includegraphics[scale=0.32]{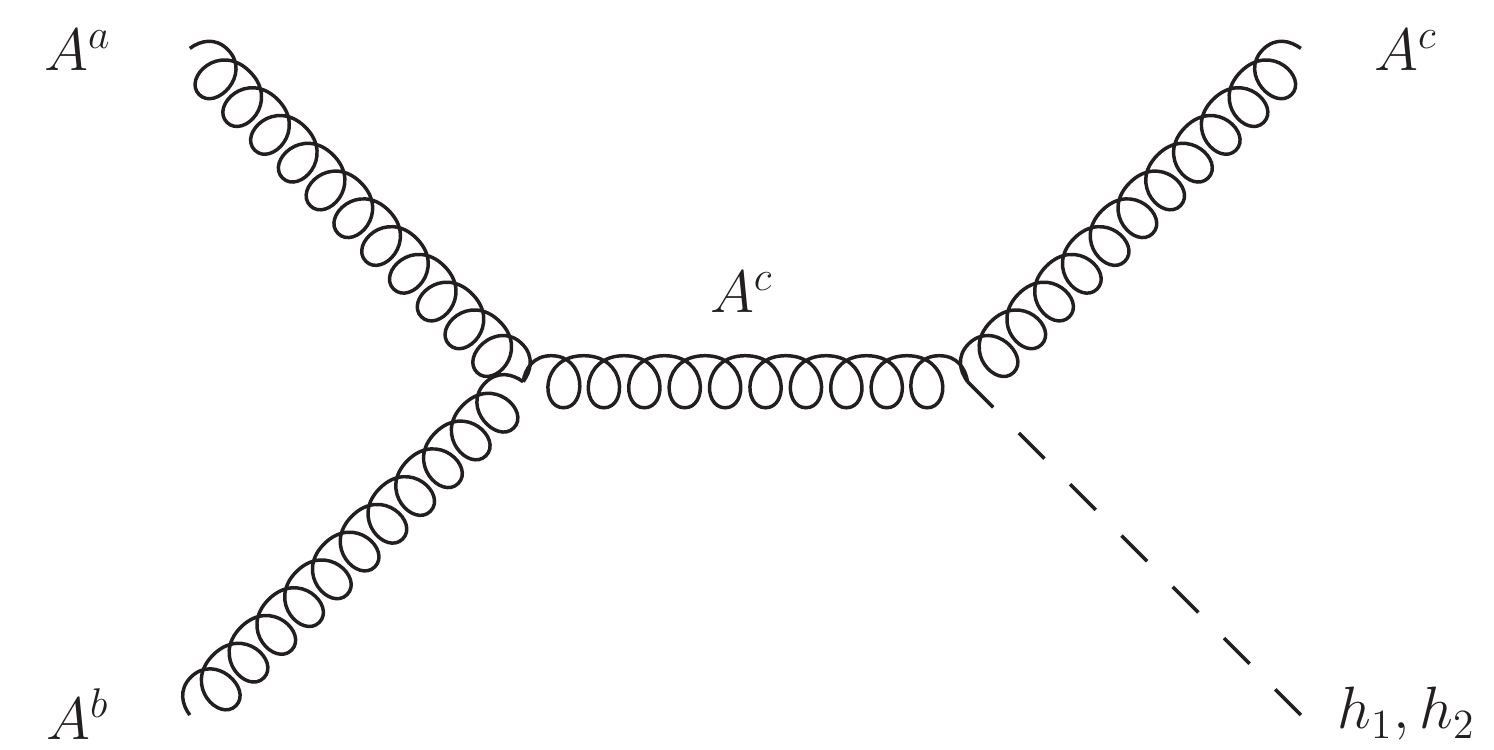} \hspace{0.8cm}
\includegraphics[scale=0.32]{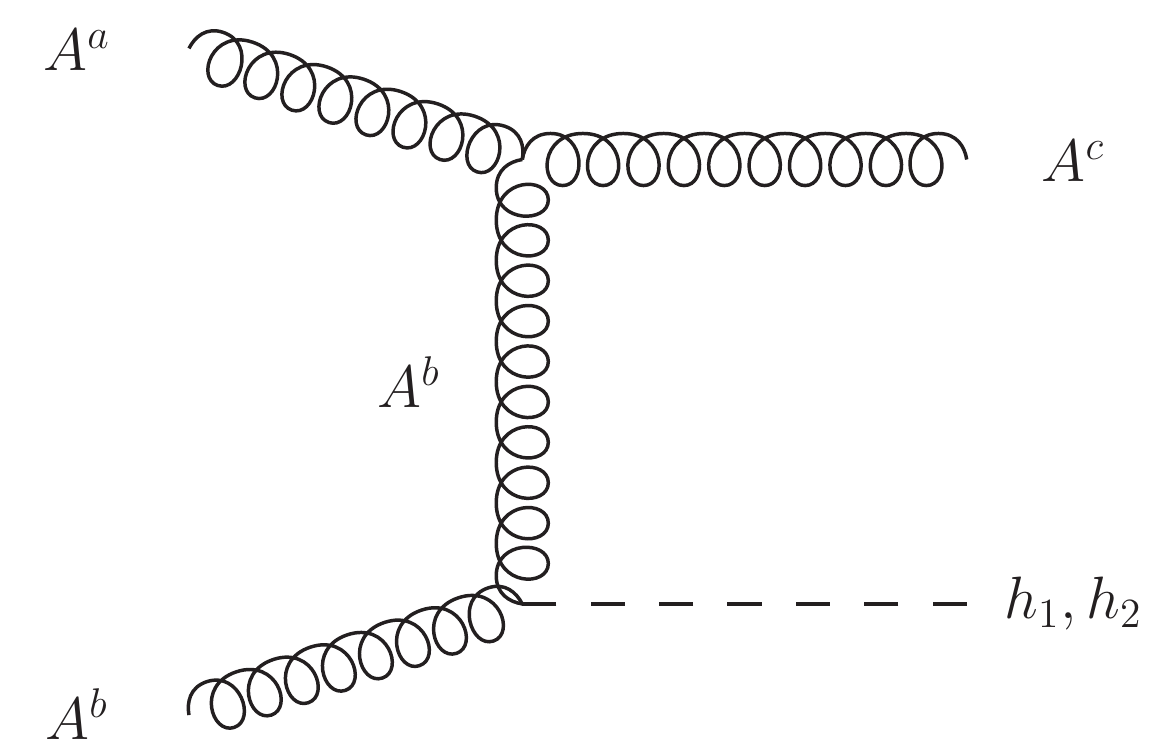}
 }
\caption{ \label{diagrams2}
Semiannihilation of vector DM.
}
\end{figure}

Non--Abelian DM features a semi--annihilation channel $AA \rightarrow A h_{1,2}$
(Fig.~\ref{diagrams2}).
In some regimes, for example at large $\tilde g$ and small $\sin\theta$, it can even dominate~\cite{D'Eramo:2012rr} (see also~\cite{Arina:2009uq,Khoze:2014xha}). 
Using the analytical results of~\cite{D'Eramo:2012rr},
we find that DM semi--annihilation is insignificant in the relevant parameter regions.
For example, at $m_A > m_{h_2}$ the $\sin\theta$--unsuppressed and potentially important channel $AA \rightarrow A h_2$ opens up, yet it is dominated by $AA \rightarrow h_2 h_2$.
Also, around the resonances the gauge coupling is rather small which diminishes the relative importance of semi--annihilation. 

No firm conclusion can be reached as to whether the region $m_A > m_{h_2}$ is allowed.
As stated above, the uncertainties in $f_N$ play a critical role due to the proximity of the Planck band and the LUX bound.

%=========================================================================
\subsection{SU(3) dark matter}
%=========================================================================
In the SU(3) case, DM is composed again of 3 species with two of them being degenerate 
($A^1_\mu, A^2_\mu$) in mass and the third one being lighter ($A^{3\prime}_\mu$). This is a result of the mixing between the gauge bosons corresponding to the Cartan generators of SU(3). Therefore, while the couplings of $A^1_\mu, A^2_\mu$ to the Higgs like scalars remain the same
as in the SU(2) case, the coupling of $A^{3\prime}_\mu$ changes. In terms of
$m_A \equiv m_{A^{1,2}}$, the relevant Lagrangian reads
\bal
\Lcal & \supset \frac12 m_{A^{}}^2 \Big(\sum_{a=1,2} A^a_\mu A^{a\mu}+\Big(1-{\tan\alpha \over \sqrt{3}}\Big) A'^3_\mu A'^{3\mu}\Big)
\\
&+ \frac{ \tilde g \, m_{A^{}}}2 (- h_1 \sin\theta + h_2 \cos\theta) \Big(\sum_{a=1,2} A^a_\mu A^{a\mu}+\Big(\cos\alpha-{\sin\alpha\over \sqrt{3}}\Big)^2 A'^3_\mu A'^{3\mu}\Big)
\nn
\\
&+ \frac{ \tilde g^2}8 (h_1^2 \sin^2\theta -2 h_1 h_2 \sin\theta \cos\theta + h_2^2 \cos^2 \theta) \Big( \! \sum_{a=1,2} \!\! A^a_\mu A^{a\mu} \! + \! \Big(\cos \alpha-{\sin \alpha \over\sqrt{3}}\Big)^2 A'^3_\mu A'^{3\mu}\Big) .
\nn
\eal
The mass of the lighter DM component is reduced by the factor 
$( 1- {1\over \sqrt{3}} \tan\alpha )^{1/2} $ compared to that of 
$A^1_\mu$ and $A^2_\mu$, while the gauge--scalar couplings decrease by a factor
$( \cos\alpha- {1\over \sqrt{3}} \sin\alpha )^{ 2}$.
Therefore, the lighter state has a smaller annihilation cross section. 
Note that 
$\sin \alpha \simeq \frac{\sqrt{3} v_2^2}{4 v_1^2}$ and even a factor 
of two difference in the triplet VEVs leads to a rather small $\alpha \sim 10^{-1}$.
In that case, there is no tangible difference between the SU(2) and the SU(3) analyses.
\begin{figure}[t] 
\centering{
\includegraphics[scale=0.6]{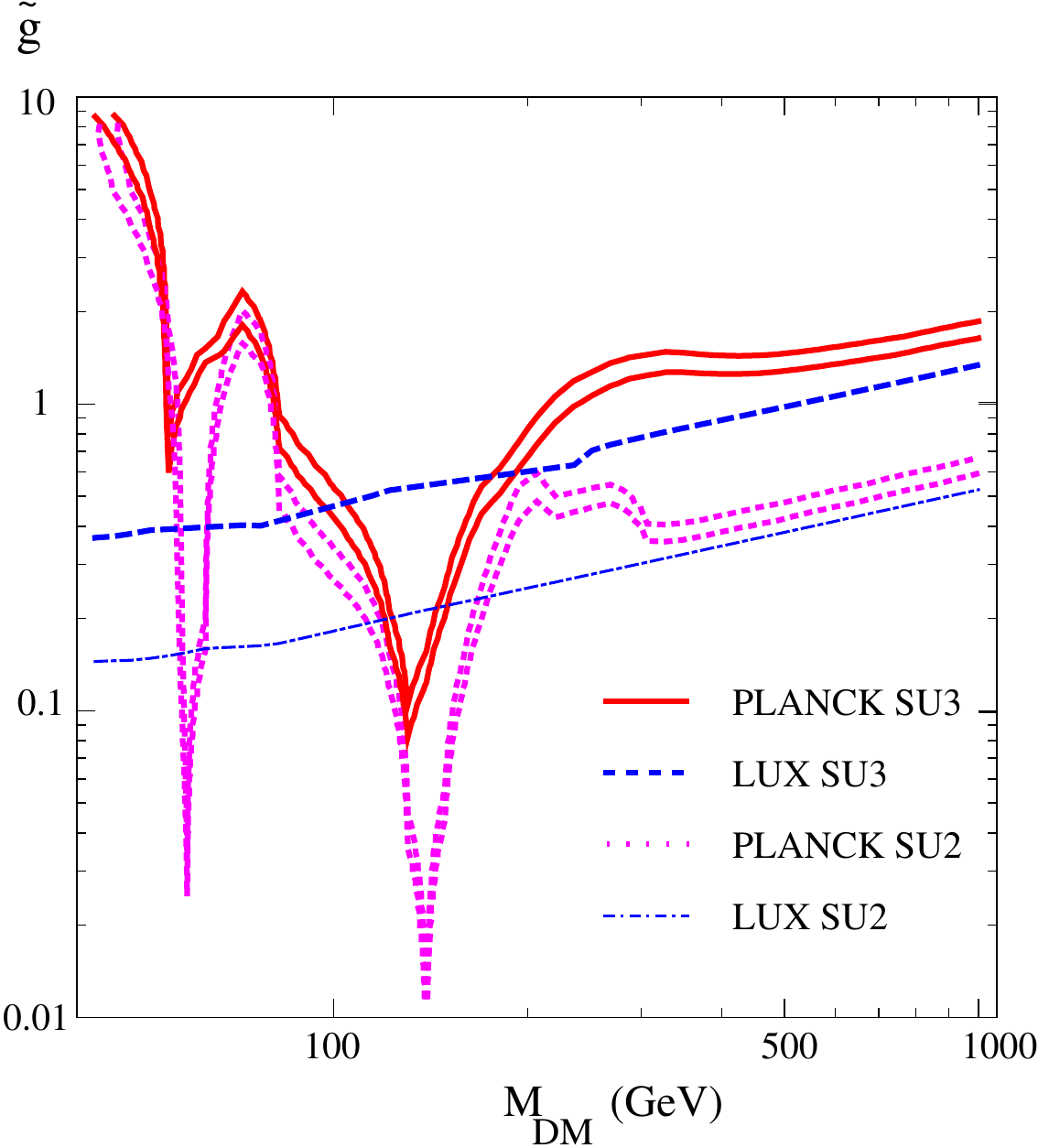}
}
\caption{ \label{plot2}
 Illustration of the main features of SU(3) vs SU(2) DM in the $extreme$ case. Here $v_1/v_2 = 1.1$ and $\sin\theta=0.3, m_{h_2}= 280$ GeV. In the SU(3) case, $M_{\rm DM}$ stands for the mass of the (dominant) lighter component $A_\mu^{3\prime}$.
 See Fig.~\ref{plots} for further details. 
 \label{su3plot}
}
\end{figure}

To understand where differences can appear, it is instructive to consider the limit $v_1 \simeq v_2$. Although in this case there are further relatively light states
that can mediate DM annihilation (e.g. $A_\mu^{8\prime}$), let us consider the simplified example in which only the same states are allowed to contribute in the SU(2) and SU(3) 
set--ups. The main features (Fig.~\ref{su3plot}) are that the gauge coupling must be larger in the SU(3) case
in order to allow for efficient annihilation of $A^{3\prime}_\mu$ and that the resonant
dips are slightly shifted due to a different freeze--out temperature (see e.g.~\cite{Griest:1990kh}). The DM density today is dominated by the lighter component. Since it couples to nucleons weaker than $A^{1,2}_\mu$ do, the direct detection constraint relaxes. Understanding further features of the model
would require precise knowledge of the spectrum and the couplings, which we
relegate to future work~\cite{nextpaper}.

Finally, one should keep in mind that 
 there exists the coupling $A^{3\prime}_\mu \; \tilde\varphi_3 \;\tilde\varphi_4$,
where $A^{3\prime}$ and $\tilde\varphi_4$ have the same parities.
If $\tilde\varphi_4$
were light, that is, $\lambda_4-\lambda_5 \ll 1$ and/or $v_1 \sim v_2$ (see Eq.~(\ref{phi4-mass})), 
the decay $A^{3\prime}_\mu \rightarrow \tilde\varphi_4 \, +$~SM would occur. In that case, DM would consist of both the vector and scalar components.
We defer a detailed study of this scenario to future work~\cite{nextpaper}.

%=========================================================================
%=========================================================================
\section{Summary and conclusions}
%=========================================================================
%=========================================================================
In this paper, we have considered the possibility that the hidden sector enjoys SU(N) gauge symmetry and couples to the Standard Model through the Higgs portal. We find that when endowed with a ``minimal'' matter content, such hidden sectors lead naturally to stable vector dark matter. The underlying Lie group symmetries which stabilise DM 
are associated with complex conjugation of the group elements and discrete gauge transformations.

We require complete breaking of hidden SU(N) by scalar multiplets to avoid massless states (barring confinement in some cases). That can be done in a minimal fashion by introducing $N-1$ scalar multiplets in the fundamental representation, which develop generic VEVs.
If the scalar sector preserves {\it CP}, the above--mentioned discrete symmetries of the Lie group 
generalise to full--fledged symmetries of the model and lead to stable gauge fields.
When sufficiently light, they constitute all of dark matter. 
In this case, DM consists of 3 components 
associated with an SU(2) subgroup which hosts the lightest gauge fields $A_\mu^{1\prime}$, $A_\mu^{2\prime}$ and $A_\mu^{3\prime}$. Two of them ($A_\mu^{1\prime}, A_\mu^{2\prime}$) are
always degenerate in mass, while for $N=2$ all 3 components have the same mass.

We have performed phenomenological analyses of U(1), SU(2) and SU(3) gauge field dark matter. We find that there are vast regions of parameter space where all of the relevant constraints are satisfied. In many of these regions, DM annihilation is facilitated by 
the broad resonances associated with the Higgs--like scalars.
We also find that the SU(3) case appears very similar to that of SU(2), 
unless the scalar VEVs breaking SU(3) are close in magnitude.

\vspace{10pt}
{\bf Acknowledgements.} 
 The authors are indebted to Sasha Pukhov and Genevieve Belanger for their
help with the new version of micrOMEGAs 4.1.8. C.G. and O.L. acknowledge support
from the Academy of Finland, project ``The Higgs boson and the Cosmos''.

This work was also supported by the Spanish MICINN's Consolider-Ingenio 2010
Programme under grant Multi-Dark {\bf CSD2009-00064}, the contract
{\bf FPA2010-17747} and the France-US PICS no. 06482. 
Y.M. acknowledges
partial support from the European Union FP7 ITN INVISIBLES (Marie Curie
Actions, PITN- GA-2011- 289442) and the ERC advanced grants
 Higgs@LHC and MassTeV. This
research was also supported in part by the Research Executive Agency
(REA) of the European Union under the Grant Agreement PITN-GA2012-316704
(``HiggsTools"). 

The authors would like to thank the Instituto de Fisica Teorica (IFT
UAM-CSIC) in Madrid for its support via the Centro de Excelencia Severo
Ochoa Program under Grant SEV-2012-0249, during the Program
``Identification of Dark Matter with a Cross-Disciplinary Approach'' where
some of the ideas presented in this paper were developed.

\begin{appendix}
\section*{Appendix}

%=========================================================================
%=========================================================================
\section{Hidden SU(3) vector--scalar couplings for $v_3=0$}
%=========================================================================
%=========================================================================
The tables below provide a list of most important gauge--scalar couplings. These
are relevant to DM phenomenology as well as to understanding the decay channels of the heavier gauge fields.

 \begin{table}[h]
 \begin{center}
 \begin{tabular}{|c|c||c|c||c|}
\hline
a&b&i&j&coeff. of $\tilde A_\mu^a \tilde A^{\mu b} \tilde \varphi^i \tilde \varphi^j$
\\
\hline \hline
$4$&$4$&$1$&$1$& \multirow{4}{*}{$\tilde g^2/8$}
\\
$5$&$5$&$1$&$1$&
\\
$6$&$6$&$1$&$1$&
\\
$7$&$7$&$1$&$1$&
\\
\hline
$8$&$8$&$1$&$1$& $\tilde g^2/6$
\\
\hline \hline
$1$&$1$&$2$&$2$&\multirow{5}{*}{$\tilde g^2/8$}
\\
$2$&$2$&$2$&$2$&
\\
$3$&$3$&$2$&$2$& 
\\
$6$&$6$&$2$&$2$& 
\\
$7$&$7$&$2$&$2$& 
\\
\hline
$8$&$8$&$2$&$2$&$\tilde g^2/24$
\\
\hline
$3$&$8$&$2$&$2$&$- \frac{1}{4 \sqrt{3}}\tilde g^2$
\\
\hline 
\end{tabular}
\quad
 \begin{tabular}{|c|c||c|c||c|}
\hline
a&b&i&j&coeff. of $\tilde A_\mu^a \tilde A^{\mu b} \tilde \varphi^i \tilde \varphi^j$
\\
\hline \hline
$4$&$4$&$3$&$3$& \multirow{4}{*}{$\frac 18 \tilde g^2 \frac{v_1^2+v_2^2}{v_1^2}$}
\\
$5$&$5$&$3$&$3$&
\\
$6$&$6$&$3$&$3$&
\\
$7$&$7$&$3$&$3$&
\\
\hline
$8$&$8$&$3$&$3$& $\frac16 \tilde g^2 \frac{v_1^2+v_2^2}{v_1^2}$
\\
\hline \hline
$4$&$4$&$4$&$4$& \multirow{4}{*}{$\frac18 \tilde g^2 \frac{v_1^2+v_2^2}{v_1^2}$}
\\
$5$&$5$&$4$&$4$&
\\
$6$&$6$&$4$&$4$&
\\
$7$&$7$&$4$&$4$&
\\
\hline
$8$&$8$&$4$&$4$& $\frac16 \tilde g^2 \frac{v_1^2+v_2^2}{v_1^2}$
\\
\hline \hline
$1$&$4$&$2$&$3$& \multirow{2}{*}{$\frac14 \tilde g^2 \frac{\sqrt{v_1^2+v_2^2}}{v_1}$}
\\
$2$&$5$&$2$&$3$&
\\
\hline
$3$&$6$&$2$&$3$&$-\frac14 \tilde g^2 \frac{\sqrt{v_1^2+v_2^2}}{v_1}$
\\
\hline
$6$&$8$&$2$&$3$& $- \frac{1}{4 \sqrt{3}}\tilde g^2 \frac{\sqrt{v_1^2+v_2^2}}{v_1}$
\\
\hline \hline
$1$&$5$&$2$&$4$& $\frac14 \tilde g^2 \frac{\sqrt{v_1^2+v_2^2}}{v_1}$
\\
\hline
$2$&$4$&$2$&$4$&\multirow{2}{*}{$-\frac14 \tilde g^2 \frac{\sqrt{v_1^2+v_2^2}}{v_1}$}
\\
$3$&$7$&$2$&$4$&
\\
\hline
$7$&$8$&$2$&$4$& $- \frac{1}{4 \sqrt{3}}\tilde g^2 \frac{\sqrt{v_1^2+v_2^2}}{v_1}$
\\
\hline 
\end{tabular}
\end{center}
\caption{\label{AAscalarscalar} Non--derivative couplings $\tilde A_\mu^a \tilde A^{\mu b} \tilde \varphi^i \tilde \varphi^j$.
}
 \end{table}
 \begin{table}[h]
 \begin{center}
 \begin{tabular}{|c|c||c||c|}
\hline
a&b&i&coeff. of $\tilde A_\mu^a \tilde A^{\mu b} \tilde \varphi^i$
\\
\hline \hline
$4$&$4$&$1$& \multirow{4}{*}{$\frac{1}{4}\tilde g^2 v_1$}
\\
$5$&$5$&$1$& 
\\
$6$&$6$&$1$& 
\\
$7$&$7$&$1$& 
\\
\hline
$8$&$8$&$1$& $\frac{1}{3}\tilde g^2 v_1$
\\
\hline \hline
$1$&$1$&$2$& \multirow{5}{*}{$\frac{1}{4}\tilde g^2 v_2 $}
\\
$2$&$2$&$2$& 
\\
$3$&$3$&$2$& 
\\
$6$&$6$&$2$& 
\\
$7$&$7$&$2$& 
\\
\hline
$8$&$8$&$2$& $\frac{1}{12}\tilde g^2 v_2 $
\\
\hline
$3$&$8$&$2$& $- \frac{1}{2 \sqrt{3}}\tilde g^2 v_2 $
\\
\hline 
\end{tabular}
\quad
 \begin{tabular}{|c|c||c||c|}
\hline
a&b&i&coeff. of $\tilde A_\mu^a \tilde A^{\mu b} \tilde \varphi^i$
\\
\hline \hline
$1$&$4$&$3$& \multirow{2}{*}{$\frac{1}{4}\tilde g^2 \frac{v_2 \sqrt{v_1^2+v_2^2} }{v_1} $}
\\
$2$&$5$&$3$& 
\\
\hline
$3$&$6$&$3$& $-\frac{1}{4}\tilde g^2 \frac{v_2 \sqrt{v_1^2+v_2^2} }{v_1} $
\\
\hline
$6$&$8$&$3$& $- \frac{1}{4 \sqrt{3}}\tilde g^2 \frac{v_2 \sqrt{v_1^2+v_2^2} }{v_1} $
\\
\hline 
\hline
$1$&$5$&$4$& $\frac{1}{4}\tilde g^2 \frac{v_2 \sqrt{v_1^2+v_2^2} }{v_1} $
\\
\hline
$2$&$4$&$4$& \multirow{2}{*}{$-\frac{1}{4}\tilde g^2 \frac{v_2 \sqrt{v_1^2+v_2^2} }{v_1} $}
\\
$3$&$7$&$4$& 
\\
\hline
$7$&$8$&$4$& $- \frac{1}{4 \sqrt{3}}\tilde g^2 \frac{v_2 \sqrt{v_1^2+v_2^2} }{v_1} $
\\
\hline 
\end{tabular}
\end{center}
\caption{\label{AAscalar} Non--derivative couplings $\tilde A_\mu^a \tilde A^{\mu b} \tilde \varphi^i$. 
}
 \end{table}

\end{appendix}

{}

\end{document}